\documentclass[pre,twocolumn,amsmath,amssymb,floatfix,superscriptaddress,showpacs]{revtex4}

\usepackage{graphicx}
\usepackage{latexsym}
\usepackage{amsmath}
\usepackage{amssymb}
\usepackage{amsfonts}
\usepackage{bm}

\newcommand{\la}{\left<}
\newcommand{\ra}{\right>}
\newcommand{\fvec}{\ensuremath{\underline{f}}}

\newcommand{\nvecl}{\underline{n}_l}

\newcommand{\rvec}{\ensuremath{\underline{r}}}
\newcommand{\rvecl}{\ensuremath{\underline{r}_l}}

\newcommand{\vvec}{\ensuremath{\underline{v}}}

\newcommand{\ddiff}{\ensuremath{\text{d}}}

\newcommand{\nlx}{n_{l,x}}
\newcommand{\nly}{n_{l,y}}
\newcommand{\ul}{u_l}
\newcommand{\rl}{r_l}
\newcommand{\rx}{r_x}
\newcommand{\ry}{r_y}
\newcommand{\vx}{v_x}
\newcommand{\vy}{v_y}
\newcommand{\vix}{v_{i,x}}
\newcommand{\viy}{v_{i,y}}

\newcommand{\kB}{\mbox{$k_{\rm B}$}}

\newcommand{\NVgT}{\ensuremath{\text{NV}\gamma\text{T}}}
\newcommand{\NVtT}{\ensuremath{\text{NV}\tau\text{T}}}

\newcommand{\Pid}{\ensuremath{P_\mathrm{id}}}
\newcommand{\Pex}{\ensuremath{P_\mathrm{ex}}}

\newcommand{\muA}{\ensuremath{\mu_\mathrm{A}}}
\newcommand{\muAid}{\ensuremath{\mu_\mathrm{A,id}}}
\newcommand{\muAex}{\ensuremath{\mu_\mathrm{A,ex}}}
\newcommand{\muAhat}{\ensuremath{\hat{\mu}_\mathrm{A}}}
\newcommand{\muAidhat}{\ensuremath{\hat{\mu}_\mathrm{A,id}}}
\newcommand{\muAexhat}{\ensuremath{\hat{\mu}_\mathrm{A,ex}}}
\newcommand{\muB}{\ensuremath{\mu_\mathrm{B}}}

\newcommand{\Ctlam}{\ensuremath{\left.C(t)\right|_{\lambda}}}
\newcommand{\Meq}{M_\mathrm{eq}}

\newcommand{\Geq}{G_\mathrm{eq}}

\newcommand{\tauzero}{\tau_0}
\newcommand{\tauhat}{\hat{\tau}}

\newcommand{\tauid}{\tau_\mathrm{id}}
\newcommand{\tauidhat}{\hat{\tau}_\mathrm{id}}
\newcommand{\tauexhat}{\hat{\tau}_\mathrm{ex}}
\newcommand{\gamhat}{\hat{\gamma}}

\newcommand{\Ihat}{\hat{I}}
\newcommand{\Xhat}{\hat{X}}

\newcommand{\Acal}{\ensuremath{{\cal A}}}
\newcommand{\Bcal}{\ensuremath{{\cal B}}}
\newcommand{\ahat}{\ensuremath{\hat{a}}}
\newcommand{\bhat}{\ensuremath{\hat{b}}}
\newcommand{\Ahat}{\ensuremath{\hat{A}}}
\newcommand{\Bhat}{\ensuremath{\hat{B}}}
\newcommand{\dAhat}{\ensuremath{\delta\hat{A}}}
\newcommand{\dBhat}{\ensuremath{\delta\hat{B}}}
\newcommand{\Hhat}{\hat{\cal H}}
\newcommand{\Hidhat}{\hat{\cal H}_\mathrm{id}}
\newcommand{\Hexhat}{\hat{\cal H}_\mathrm{ex}}

\newcommand{\dtMD}{\delta t_\mathrm{MD}}
\newcommand{\fDebye}{f_\mathrm{Debye}}

\newcommand{\Geff}{G_\mathrm{eff}}
\newcommand{\Gext}{G_\mathrm{ext}}
\newcommand{\Mext}{M_\mathrm{ext}}
\newcommand{\Uext}{U_\mathrm{ext}}
\newcommand{\tauext}{\tau_\mathrm{ext}}
\newcommand{\gamext}{\gamma_\mathrm{ext}}
\newcommand{\gamzero}{\gamma_0}

\newcommand{\gammamax}{\delta \gamma_\mathrm{max}}

\newcommand{\muAB}{\ensuremath{\mu_\mathrm{ab}}}
\newcommand{\muGG}{\ensuremath{\mu_\mathrm{\gamma\gamma}}}
\newcommand{\muGT}{\ensuremath{\mu_\mathrm{\gamma\tau}}}
\newcommand{\muTT}{\ensuremath{\mu_\mathrm{\tau\tau}}}
\newcommand{\muTTid}{\ensuremath{\mu_\mathrm{\tau\tau,id}}}

\newcommand{\muTTlam}{\ensuremath{\left.\mu_\mathrm{\tau\tau}\right|_{\lambda}}}
\newcommand{\muABn}{\ensuremath{\overline{\mu}_\mathrm{ab}(n)}}
\newcommand{\muABt}{\ensuremath{\overline{\mu}_\mathrm{ab}(t)}}
\newcommand{\muGGt}{\ensuremath{\overline{\mu}_\mathrm{\gamma\gamma}(t)}}
\newcommand{\muGTt}{\ensuremath{\overline{\mu}_\mathrm{\gamma\tau}(t)}}
\newcommand{\muTTt}{\ensuremath{\overline{\mu}_\mathrm{\tau\tau}(t)}}
\newcommand{\muTTtlam}{\ensuremath{\left.\overline{\mu}_\mathrm{\tau\tau}(t)\right|_{\lambda}}}
\newcommand{\thetaAB}{\ensuremath{\theta_\mathrm{ab}}}
\newcommand{\thetaGG}{\ensuremath{\theta_\mathrm{\gamma\gamma}}}
\newcommand{\thetaGT}{\ensuremath{\theta_\mathrm{\gamma\tau}}}
\newcommand{\thetaTT}{\ensuremath{\theta_\mathrm{\tau\tau}}}
\newcommand{\gBAt}{\ensuremath{g_\mathrm{ba}(t)}}
\newcommand{\gABt}{\ensuremath{g_\mathrm{ab}(t)}}
\newcommand{\gGGt}{\ensuremath{g_\mathrm{\gamma\gamma}(t)}}
\newcommand{\gGTt}{\ensuremath{g_\mathrm{\gamma\tau}(t)}}
\newcommand{\gTTt}{\ensuremath{g_\mathrm{\tau\tau}(t)}}

\newcommand{\cABt}{\ensuremath{C_\mathrm{ab}(t)}}
\newcommand{\cBAt}{\ensuremath{C_\mathrm{ba}(t)}}
\newcommand{\cGGt}{\ensuremath{C_\mathrm{\gamma\gamma}(t)}}
\newcommand{\cGTt}{\ensuremath{C_\mathrm{\gamma\tau}(t)}}
\newcommand{\cTTt}{\ensuremath{C_\mathrm{\tau\tau}(t)}}

\newcommand{\cTTtlam}{\ensuremath{\left.C_\mathrm{\tau\tau}(t)\right|_{\lambda}}}
\newcommand{\Ggg}{G_\mathrm{\gamma\gamma}}
\newcommand{\Ggt}{G_\mathrm{\gamma\tau}}
\newcommand{\Gtt}{G_\mathrm{\tau\tau}}

\newcommand{\Gggt}{G_\mathrm{\gamma\gamma}(t)}
\newcommand{\Ggtt}{G_\mathrm{\gamma\tau}(t)}
\newcommand{\Gttt}{G_\mathrm{\tau\tau}(t)}

\newcommand{\ttraj}{t_\mathrm{traj}}
\newcommand{\tauA}{\tau_\mathrm{A}}
\newcommand{\tstar}{\tau_{\star}}
\newcommand{\thetatilde}{\tilde{\theta}}

\newcommand{\Dab}{D_\mathrm{ab}}
\newcommand{\Tab}{T_\mathrm{ab}}
\newcommand{\Dgg}{D_\mathrm{\gamma\gamma}}
\newcommand{\Tgg}{T_\mathrm{\gamma\gamma}}
\newcommand{\Dgt}{D_\mathrm{\gamma\tau}}
\newcommand{\Tgt}{T_\mathrm{\gamma\tau}}
\newcommand{\Dtt}{D_\mathrm{\tau\tau}}
\newcommand{\Ttt}{T_\mathrm{\tau\tau}}

\newcommand{\Gstor}{G^{\prime}(\omega)}
\newcommand{\Gloss}{G^{\prime\prime}(\omega)}

\begin{document}

\title{Shear-strain and shear-stress fluctuations in generalized\\
       Gaussian ensemble simulations of isotropic elastic networks}

\author{J.P.~Wittmer}
\email{joachim.wittmer@ics-cnrs.unistra.fr}
\affiliation{Institut Charles Sadron, Universit\'e de Strasbourg \& CNRS, 23 rue du Loess, 67034 Strasbourg Cedex, France}
\author{I. Kriuchevskyi}
\affiliation{Institut Charles Sadron, Universit\'e de Strasbourg \& CNRS, 23 rue du Loess, 67034 Strasbourg Cedex, France}
\author{J. Baschnagel}
\affiliation{Institut Charles Sadron, Universit\'e de Strasbourg \& CNRS, 23 rue du Loess, 67034 Strasbourg Cedex, France}
\author{H.~Xu}
\affiliation{LCP-A2MC, Institut Jean Barriol, Universit\'e de Lorraine \& CNRS, 1 bd Arago, 57078 Metz Cedex 03, France}

\begin{abstract}
Shear-strain and shear-stress correlations in isotropic elastic bodies are investigated both
theoretically and numerically at either imposed mean shear-stress $\tau$ ($\lambda=0$) or 
shear-strain $\gamma$ ($\lambda=1$) and for more general values of a dimensionless parameter 
$\lambda$ characterizing the generalized Gaussian ensemble. It allows to tune the strain fluctuations 
$\muGG \equiv \beta V \la \delta \gamhat^2 \ra = (1-\lambda)/\Geq$ with 
$\beta$ being the inverse temperature, $V$ the volume, $\gamhat$ the instantaneous strain
and $\Geq$ the equilibrium shear modulus.
Focusing on spring networks in two dimensions we show, e.g., for the stress 
fluctuations $\muTT \equiv \beta V \la \delta \tauhat^2 \ra$
($\tauhat$ being the instantaneous stress)
that $\muTTlam = \muA - \lambda \Geq$ with $\muA =  \muTT|_{\lambda=0}$ being the 
affine shear-elasticity.
For the stress autocorrelation function
$\cTTt \equiv \beta V \la \delta \tauhat(t) \delta \tauhat(0) \ra$
this result is then seen (assuming a sufficiently slow shear-stress barostat)
to generalize to $\cTTtlam = G(t) - \lambda \Geq$
with $G(t) = \cTTt|_{\lambda= 0}$ being the shear-stress relaxation modulus.
\end{abstract}
\date{\today}
\maketitle

\begin{figure}[t]
\centerline{\resizebox{1.0\columnwidth}{!}{\includegraphics*{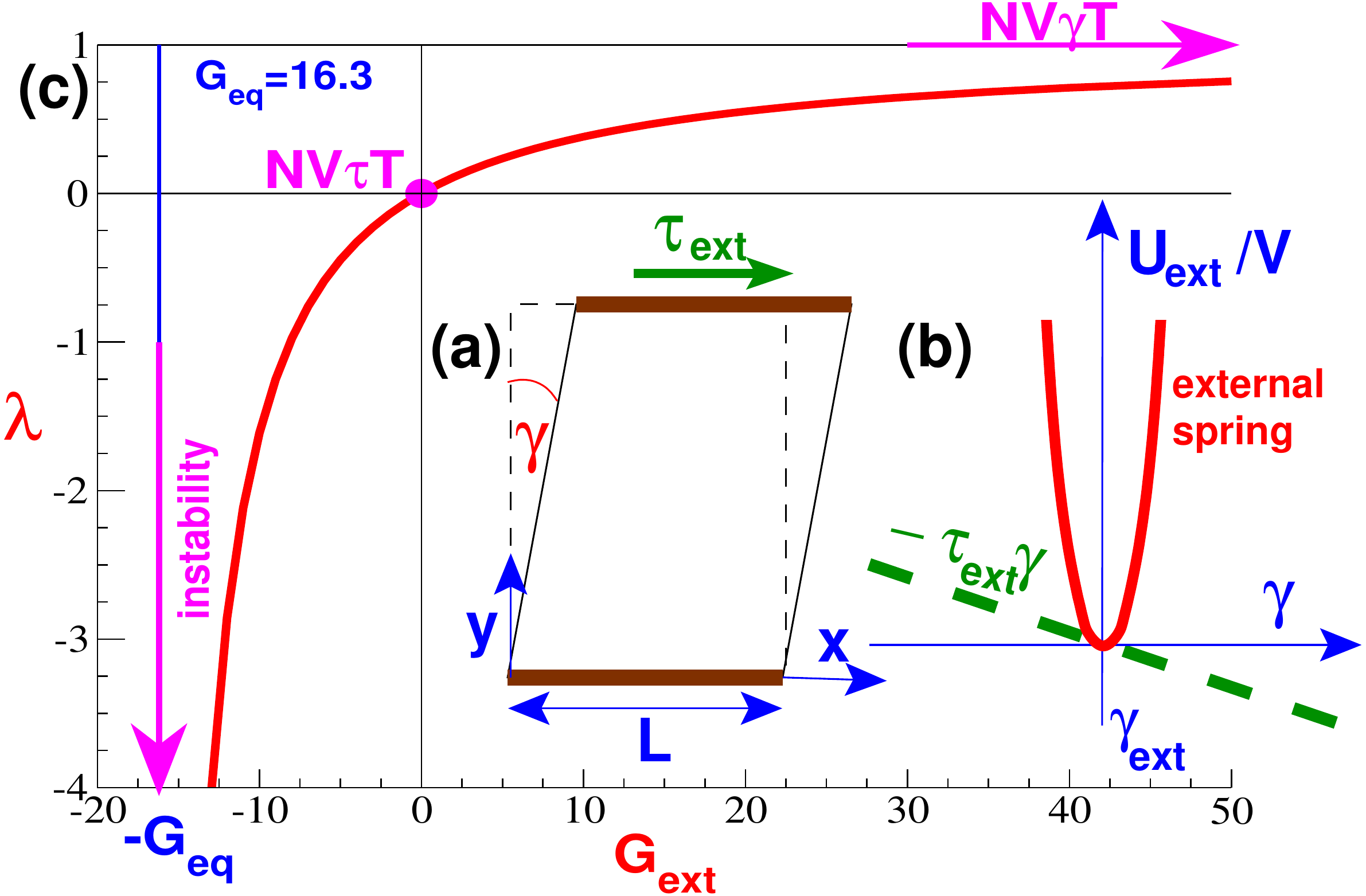}}}
\caption{Sketch of addressed problem:
{\bf (a)}
Shear strain experiment in the $(x,y)$-plane with shear strain $\gamma$
and applied external shear stress $\tauext$.
{\bf (b)}
External potential $\Uext(\gamma)/V$ according to eq.~(\ref{eq_Uext}) with the bold line indicating
the external spring of spring constant $\Gext$.
{\bf (c)} 
Dimensionless parameter $\lambda = \Gext/(\Geq+\Gext)$ comparing the
external spring constant $\Gext$ and the equilibrium shear modulus $\Geq$.
We set $\Geq=16.3$ as for the elastic model system considered below. 
\label{fig_sketch}
}
\end{figure}

\section{Introduction}
\label{sec_intro}

\paragraph*{Background.}
A ``simple average" $A = \langle \hat{A} \rangle$ of an observable ${\cal A}$ \cite{AllenTildesleyBook}
does not depend on which thermodynamic ensemble it is measured in, at least not if the system is 
sufficiently large and if each ensemble samples indeed the same thermodynamic state point
\cite{Lebowitz67,Callen,AllenTildesleyBook,FrenkelSmitBook,LandauBinderBook,ThijssenBook}. 
An example for such a simple average is the affine shear-elasticity $\muA$ characterizing the energy 
change of an affinely shear-strained elastic body \cite{BornHuang,WXP13,WXB15,WXBB15} as properly defined 
below in sect.~\ref{sec_algo}. As one may verify numerically \cite{WXP13}, it is irrelevant
whether one computes $\muA$ in the $\NVgT$-ensemble at constant particle number $N$, volume $V$, 
shear strain $\gamma$ and temperature $T \equiv 1/\beta$ or 
in the conjugated $\NVtT$-ensemble where the strain is allowed
to fluctuate subjected to the constraint that the internal mean shear stress $\tau$ is imposed by
the external applied shear stress $\tauext$ \cite{Callen}.
In contrast to this, the fluctuation $\langle \dAhat \dBhat \rangle$ of two 
observables ${\cal A}$ and ${\cal B}$ may depend on whether an extensive variable $X$ or its conjugated 
intensive variable $I$ is imposed \cite{Lebowitz67,Callen,AllenTildesleyBook}. 
Focusing in the present work on shear-strained isotropic 
elastic networks, as sketched in panel (a) of fig.~\ref{fig_sketch}, the relevant extensive 
variable is the rescaled shear strain $X=V\gamma$ and the conjugated intensive variable the 
shear stress $I=\tau$. 
Using the Lebowitz-Percus-Verlet (LPV) transform \cite{Lebowitz67} it is seen
for the (rescaled) mean-squared fluctuation $\muTT \equiv \beta V \la \delta \tauhat^2\ra$ 
of the instantaneous shear stress $\tauhat$ that \cite{WXP13,WXB15}
\begin{equation}
\muTT|_{\lambda=0} = \muTT|_{\lambda=1} + \Geq
\label{eq_muTT_trans_old}
\end{equation}
with $\Geq$ being the equilibrium shear modulus.
For later convenience the $\NVtT$-ensemble is indicated by ``$\lambda=0$"
and the $\NVgT$-ensemble by ``$\lambda=1$".
Generalizing eq.~(\ref{eq_muTT_trans_old}) in the time-domain it has been shown for the
stress-stress correlation function $\cTTt \equiv \beta V \langle \delta \tauhat(t) \delta \tauhat(0) \rangle$
that \cite{WXB15,WXBB15}
\begin{equation}
\cTTt|_{\lambda=0} = \cTTt|_{\lambda=1} + \Geq.
\label{eq_cTTt_trans_old}
\end{equation}
This assumes that the shear-barostat needed to sample the $\NVtT$-ensemble must act very slowly
on the time-scales used to probe $\cTTt|_{\lambda=0}$. This may be realized equivalently by
averaging over an ensemble of configurations with quenched strains $\gamhat$
distributed according to the $\NVtT$-ensemble \cite{WXB15}. It is due to the ergodicity breaking 
generated by the averaging over the quenched ensemble that $\cTTt|_{\lambda=0} \to \Geq$ stays
finite for $t \to \infty$, while $\cTTt|_{\lambda=1}$ must vanish for large times
(as one commonly expects for all correlation functions in ergodic systems \cite{HansenBook}).
By integration by parts it is seen that $\cTTt|_{\lambda=0}$ is equivalent to the
experimentally important shear relaxation modulus $G(t)$ \cite{WXB15}. 
The transform eq.~(\ref{eq_cTTt_trans_old})
thus implies $G(t) = \Geq + \cTTt|_{\lambda=1}$, i.e. the relaxation modulus may be measured
using the equilibrium modulus $\Geq$ and the correlation function $\cTTt|_{\lambda=1}$
both determined in the \NVgT-ensemble \cite{WXB15}.

\paragraph*{Generalized Gaussian ensemble.}
As sketched in panel (b) of fig.~\ref{fig_sketch}, it is straightforward to interpolate between
the \NVgT-ensemble and the \NVtT-ensemble by imposing an external field
\begin{equation}
\Uext(\gamhat)/V = - \tauext (\gamhat-\gamext) + \frac{1}{2} \Gext (\gamhat-\gamext)^2
\label{eq_Uext}
\end{equation}
with $\Gext$ being the spring constant of the external harmonic spring. 
The standard $\NVtT$-ensemble at $\tau = \tauext$ is recovered by setting $\Gext=0$.
Note that our approach is conceptually similar to the so-called ``Gaussian ensemble" proposed some years ago by
Hetherington and others \cite{Hetherington87,CETT06} generalizing the Boltzmann weight of the canonical
ensemble by an exponential factor $\Uext(\hat{E}) \propto \hat{E}^2$ of the instantaneous energy $\hat{E}$.
A similar external spring potential has been also used in the recent ``elastic bath" approach
by Workum and de Pablo \cite{Pablo03}.
Choosing the reference strain $\gamext$ equal to the mean strain $\gamzero$ of the $\NVtT$-system at vanishing
shear stress $\tau = \la \tauhat \ra = \tauext \equiv 0$ allows to work at constant zero mean shear stress
irrespective of the strength of the external potential \cite{foot_generalensemble}.
All the ensembles considered correspond thus to the {\em same} thermodynamic state,
i.e. all first derivatives of the energy or the free energy \cite{Callen}
and all simple averages are identical. 
As sketched in panel (c) of fig.~\ref{fig_sketch}, 
$\Gext$ is not necessarily positive, reducing the strain fluctuations,
but may even become negative, which thus amplifies the fluctuations. It has been argued \cite{Pablo03}
that this may allow a more convenient determination of the elastic modulus. 
Since the external spring is parallel to the system, the combined system and external device have an 
effective spring constant $\Geff = \Geq + \Gext$. Defining the dimensionless parameter 
$\lambda \equiv \Gext/(\Geq+ \Gext)$ the strain fluctuations of the combined system are thus given by 
\begin{equation}
\muGG \equiv \beta V \la \delta \gamhat^2 \ra = 1/\Geff = (1-\lambda)/\Geq,
\label{eq_dgam_L}
\end{equation}
i.e. must vanish linearly with $\lambda$.
\NVtT-ensemble statistics is expected for $\lambda \to 0$, 
while \NVgT-statistics should become relevant in the opposite limit for $\Gext \to \infty$, 
i.e. $\lambda \to 1$. The system must become unstable in the limit $\Gext \to -\Geq$, i.e. $\lambda \to -\infty$.

\paragraph*{Key results.}
The aim of the present study is to generalize the relations for static fluctuations, 
eq.~(\ref{eq_muTT_trans_old}), and dynamical correlation functions, eq.~(\ref{eq_cTTt_trans_old}), 
to the more general transformations between Gaussian ensembles 
characterized by the continuous parameter $\lambda \le 1$. 
In this way we want to make manifest that these transformation relations are generated 
by the {\em continuous change} of the constraint imposed on the fluctuations of the extensive variable.
Focusing on spring networks in two dimensions we show, e.g., for the stress-stress
fluctuations $\muTT$ in different $\lambda$ ensembles that
\begin{equation}
\muTT|_{\lambda} = \muTT|_{\lambda=0} - \lambda \Geq
\label{eq_keystat}
\end{equation}
with $\muTT|_{\lambda=0}$ being given by the affine shear-elasticity $\muA$ mentioned above.
Assuming a very slowly acting shear-barostat, which is irrelevant for the system evolution 
for short times $t \ll \tstar(\lambda)$, the above result is then seen to generalize in the 
time-domain for the stress-stress correlation function
\begin{equation}
\cTTt|_{\lambda} = \cTTt|_{\lambda=0} - \lambda \Geq \mbox{ for } t \ll \tstar(\lambda).
\label{eq_keydyna}
\end{equation}
Since $G(t) = \cTTt|_{\lambda=0}$, eq.~(\ref{eq_keydyna}) allows the determination
of the relaxation modulus $G(t)$ from the equilibrium modulus $\Geq$ and 
the stress-stress correlation function $\cTTt|_{\lambda}$ for any $\lambda$
and $t \ll \tstar(\lambda)$. The upper time limit $\tstar(\lambda)$ is seen 
to be set by the diffusion time $\Tgg$ of the instantaneous strain $\gamhat(t)$
over the typical strain fluctuations for $\lambda$. 
As a consequence from eq.~(\ref{eq_keystat}) and eq.~(\ref{eq_keydyna}) it appears
that it is the equilibrium modulus $\Geq$ of the system, which generates both transforms
\begin{equation}
\frac{\ddiff}{\ddiff \lambda} \left. \muTT\right|_{\lambda} = 
\frac{\ddiff}{\ddiff \lambda} \left. \cTTt\right|_{\lambda} 
= -\Geq
\label{eq_blurb}
\end{equation}
with $\left. \muTT\right|_{\lambda} = \muA$ and $\left. \cTTt\right|_{\lambda} = G(t)$
for $\lambda=0$.
Similar relations, allowing also to determine the creep compliance $J(t)$ 
\cite{WittenPincusBook,RubinsteinBook},
are obtained for strain-strain and strain-stress correlations functions.

\paragraph*{Outline.}
The stated key relations eq.~(\ref{eq_keystat}) and eq.~(\ref{eq_keydyna}) are justified 
theoretically in sect.~\ref{sec_theo}. The static fluctuations are discussed in sect.~\ref{theo_stat}
before we address the dynamical correlation functions in sects.~\ref{theo_dyna_genrel},
\ref{theo_dyna_model} and \ref{theo_dyna_lambda}  and the macroscopic linear response in sect.~\ref{theo_dyna_GtJt}.
Algorithmic details are given in sect.~\ref{sec_algo}
where the specific model system considered is introduced in sect.~\ref{algo_model}.
The canonical affine plane shear transformations used in this work are specified in
sect.~\ref{algo_affine}, the instantaneous shear stress $\tauhat$ and the
instantaneous affine shear-elasticity $\muAhat$ in sect.~\ref{algo_tau_muA}.
The zero-temperature ground state properties of the two-dimensional elastic network
are summarized in sect.~\ref{algo_refer}. 
Some technical details related to the finite-temperature simulation of the Gaussian 
$\lambda$-ensemble using a Metropolis Monte Carlo (MC) scheme as a function of the
maximum strain displacement rate $\kappa$, the second key operational parameter
of this study, are given in sect.~\ref{algo_finiteT}.
%
Section~\ref{sec_simu} presents the numerical results obtained by molecular dynamics (MD) simulations
at one finite (albeit small) temperature $T$ and one relatively
high value of the friction constant $\zeta$ of the Langevin thermostat used
\cite{AllenTildesleyBook}.
The relevant static properties are described in sect.~\ref{simu_stat}
before we turn to the dynamical strain-strain (sect.~\ref{simu_dyna_GG}),
strain-stress (sect.~\ref{simu_dyna_GT}) and stress-stress (sect.~\ref{simu_dyna_TT})
correlation functions. 
Our work is summarized in sect.~\ref{sec_conc}.

\section{Theoretical considerations}
\label{sec_theo}

\subsection{Static fluctuations}
\label{theo_stat}

\paragraph*{Lebowitz-Percus-Verlet transform.}
We begin by restating the LPV transform in a convenient form assuming that the
relevant extensive variable is the (rescaled) shear strain $X = V \gamma$
and the conjugated intensive variable the shear stress $I = \tau$.
Following Lebowitz {\em et al.} \cite{Lebowitz67,AllenTildesleyBook}, 
one verifies (see also sect. II.A of ref.~\cite{WXP13}) that to leading order 
\begin{equation}
\left. \la \dAhat \dBhat \ra\right|_{\lambda=0} = \left. \la \dAhat \dBhat \ra\right|_{\lambda=1}  
+ \frac{\Geq}{\beta V} \ \frac{\partial A}{ \partial \tau} \frac{\partial B}{ \partial \tau}
\label{eq_LPVold}
\end{equation}
for the transformation of $\langle \dAhat \dBhat \rangle$ 
with $\dAhat \equiv \Ahat - \langle \Ahat \rangle$ and $\dBhat \equiv \Bhat - \langle \Bhat \rangle$ 
from the $\NVgT$-ensemble ($\lambda=1$) to the $\NVtT$-ensemble ($\lambda=0$).
The more general transformation between arbitrary $\lambda$-ensembles can be found by
reworking the saddle-point approximation of ref.~\cite{Lebowitz67} taking into account that
the fluctuations around the peak of the distribution of the extensive variable is now
not set by the modulus $\Geq$ of the system but by the effective modulus $\Geff = \Geq + \Gext$.
How this may be done can be seen in sect.~2.5 of ref.~\cite{WXP13c} for the volume $X=V$
as extensive variable and the (negative) pressure $I=-P$ as intensive variable.
Rewriting the latter result to the present case we get the generalized LPV transform
\begin{equation}
\left. \la \dAhat \dBhat \ra\right|_{\lambda} =
\left. \la \dAhat \dBhat \ra\right|_{\lambda=1}
+ (1-\lambda) \frac{\Geq}{\beta V} \ \frac{\partial A}{ \partial \tau} \frac{\partial B}{ \partial \tau}
\label{eq_LPVgen}
\end{equation}
which reduces for $\lambda=0$ to eq.~(\ref{eq_LPVold}).

\paragraph*{Strain-strain fluctuations.}
Let us check the ensemble dependence of the rescaled strain-strain fluctuations 
$\muGG \equiv \beta V \la \delta \gamhat^2 \ra$ \cite{foot_betaV}.
The generalized Gaussian ensemble corresponds to replacing the shear modulus $\Geq$ of the system 
by the effective modulus $\Geff=\Geq+ \Gext$. 
As already stated above, eq.~(\ref{eq_dgam_L}), this leads to \cite{foot_tildefunct} 
\begin{equation}
\muGG = 1/\Geff = (1-\lambda)/\Geq \label{eq_muGG_lambda}.
\end{equation}
One verifies readily that the postulated LPV transform for general $\lambda$, eq.~(\ref{eq_LPVgen}), 
is consistent with this result.  To see this one sets $\Acal =\Bcal = V \gamma$ and uses the fact 
that the strain fluctuations must vanish in the $\NVgT$-ensemble, i.e. that the first term on the right hand-side
of eq.~(\ref{eq_LPVgen}) must vanish.

\paragraph*{Strain-stress fluctuations.}
Setting $\Acal = V \gamma$ and $\Bcal = \tau$ and using again that
the $\NVgT$-term in eq.~(\ref{eq_LPVgen}) must vanish, it is seen from the LPV transform
that the strain-stress fluctuations $\muGT \equiv \beta V \la \delta \gamhat \delta \tauhat \ra$
\cite{foot_betaV} should scale as
\begin{equation}
\muGT = 1 - \lambda.
\label{eq_muGt_lambda}
\end{equation}
This result can be also obtained directly by replacing in the definition of $\muGT$ the fluctuation 
$\delta \tauhat$ by $\Geq \delta \gamhat$ and using then the strain-strain relation, eq.~(\ref{eq_muGG_lambda}).

\paragraph*{Stress-stress fluctuations.}
We turn to the most important stress-stress fluctuation
$\muTT \equiv \beta V \la \delta \tauhat^2 \ra$ \cite{foot_betaV}. 
We set now $\Acal = \Bcal = \tau$ in the LPV transform. Since the
stress fluctuations do {\em not} vanish in the $\NVgT$-ensemble
the corresponding term must now be included. This yields \cite{foot_tildefunct}
\begin{equation}
\muTT|_{\lambda} = \muTT|_{\lambda=1} + (1-\lambda) \Geq. 
\label{eq_muTT_lambda_one}
\end{equation}
Interestingly, the contribution $(1-\lambda)\Geq$ may be rewritten using the notation 
$\langle f(\gamhat) \rangle_{\gamma} \equiv \int \ddiff \gamhat \ p(\gamhat;\lambda) f(\gamhat)$
for the strain-average of a property $f(\gamhat)$ with $p(\gamhat;\lambda)$ being the normalized 
strain-distribution for the considered $\lambda$-ensemble. 
The total mean stress $\tau$ of the ensemble is thus given by
$\tau = \langle \tau(\gamhat) \rangle_{\gamma}$ with $\tau(\gamhat)$ being
the average shear-stress of all configurations of shear-strain $\gamhat$.
Using that $p(\gamhat;\lambda)$ is a Gaussian and that
$\delta \tau(\gamhat) \equiv \tau(\gamhat) - \tau = \Geq (\gamhat - \gamzero)$
with $\gamzero$ being the maximum of the distribution, it is then seen that
\begin{equation}
(1-\lambda) \Geq = \frac{\Geq^2}{\Geff} 
= \beta V \la \delta \tau(\gamhat)^2 \ra_{\gamma}.
\label{eq_fluctu_one}
\end{equation}
Using eq.~(\ref{eq_muTT_lambda_one}) this implies in turn
\begin{equation}
\muTT|_{\lambda} = \muTT|_{\lambda=1} + \beta V \la \delta \tau(\gamhat)^2 \ra_{\gamma}
\label{eq_fluctu_two}
\end{equation}
as one expects to lowest order from a standard saddle-point approximation.

\paragraph*{Stress-fluctuation formula for shear modulus.}
Substracting the transform for $\lambda=0$ from eq.~(\ref{eq_muTT_lambda_one}) confirms 
the key result eq.~(\ref{eq_keystat}). 
As shown in ref.~\cite{WXBB15}, $\muTT|_{\lambda=0}$ can be reduced by
integration by parts to the affine shear-elasticity $\muA$. Since the latter
expression is a simple average, i.e. the same value $\muA$ is obtained in any ensemble,
eq.~(\ref{eq_keystat}) may be further simplified as \cite{foot_tildefunct}
\begin{equation}
\muTT|_{\lambda} = \muA -\lambda \Geq.  
\label{eq_muTT_lambda_final}
\end{equation}
It is thus sufficient to compute the material constants $\muA$ and $\Geq$ in
any ensemble to obtain the stress-stress fluctuations $\muTT$ as a function of $\lambda$.
Note that the special case for $\lambda=1$ 
corresponds to the well-known stress-fluctuation formula 
\cite{FrenkelSmitBook,Hoover69,Barrat13,WXP13}
\begin{equation}
\Geq = \muA - \muTT|_{\lambda=1}
\label{eq_Gtt}
\end{equation}
used in numerous numerical studies to compute the modulus $\Geq$ conveniently in the $\NVgT$-ensemble
\cite{Barrat88,WTBL02,TWLB02,WXP13,WXP13c,Barrat13,Szamel15,WXB15,WXBB15}.

\subsection{Dynamics: Definitions and general relations}
\label{theo_dyna_genrel}

\paragraph*{Introduction.}
We define now several dynamical observables of interest,
which will be investigated numerically in sect.~\ref{sec_simu}, and remind 
some well-known general relations \cite{HansenBook,DoiEdwardsBook,AllenTildesleyBook}.
Time translational symmetry $(t \leftrightarrow t + \delta t)$ and 
time reversal symmetry $(t \leftrightarrow -t)$ are assumed.
We use $\ahat(t)$ and $\bhat(t)$ for either the instantaneous 
shear strain $\gamhat$ or the shear stress $\tauhat$,
$a \equiv \langle \ahat \rangle$ and $b \equiv \langle \bhat \rangle$ for
their thermodynamic averages and $\delta \ahat(t) \equiv \ahat(t) - a$ and
$\delta \bhat(t) \equiv \bhat(t) - b$ for their time-dependent fluctuations.

\paragraph*{Mean-square displacements.}
We define (generalized) mean-square displacements (MSD) by 
\begin{equation}
\gABt \equiv \frac{\beta V}{2} \la (\ahat(t) - \ahat(0)) (\bhat(t) - \bhat(0)) \ra
\label{eq_gABt_def}
\end{equation}
where the prefactor $\beta V/2$ has been introduced for convenience \cite{foot_betaV}.
Obviously, $\gABt = \gBAt$, $g_\mathrm{ab}(t)=g_\mathrm{ab}(-t)$
as may be seen using the above-mentioned symmetries, $\gABt \to 0$ for $t \to 0$  and 
\begin{equation}
\gABt \to \muAB \equiv \beta V \la \delta \ahat \ \delta \bhat \ra \mbox{ for } t \gg \Tab
\label{eq_gABt_long}
\end{equation}
with $\Tab$ being the characteristic time needed to reach this thermodynamic limit \cite{foot_ergodic}. 

\paragraph*{Correlation functions.}
Similarly, we define dynamic correlation functions by 
$\cABt \equiv \beta V \langle \delta \ahat(t) \delta \bhat(0) \rangle$ \cite{foot_betaV}.
Obviously, $C_\mathrm{ab}(0)=\muAB$, $C_\mathrm{ab}(t)=C_\mathrm{ab}(-t)$ and $\cABt = \cBAt$. 
The latter identity is seen from \cite{HansenBook,Callen}
\begin{equation}
\la \ahat(t) \bhat(0) \ra = \la \ahat(0) \bhat(-t) \ra = \la \bhat(t) \ahat(0) \ra
\label{eq_cABt2cBAt}
\end{equation}
where the time translational invariance is used in the first step
and the time reversal symmetry in the second step.
Using again both symmetries one verifies that \cite{HansenBook,DoiEdwardsBook}
\begin{equation}
\gABt = C_\mathrm{ab}(0) - \cABt = \muAB - \cABt.
\label{eq_gABt_cABt}
\end{equation}
This implies that $\cABt \to 0$ for large times $t \gg \Tab$ \cite{foot_ergodic}.
Albeit $\gABt$ and $\cABt$ thus contain the same information it will be sometimes 
better for theoretical or numerical reasons to focus on either $\gABt$ or $\cABt$.
Since $\gABt \to 0$ for $t \to 0$ it will be natural, e.g., to consider the 
double-logarithmic representation of $\gABt$ to clarify the power-law scaling
of the correlations at short times.

\paragraph*{Finite-time dependence of fluctuation estimation.}
As defined in sect.~\ref{theo_stat}, $\muAB$ is 
a thermodynamic ensemble average, hence, a time-independent property.
In practice, $\muAB$ may, however, often be estimated using a time series $(a_k,b_k)$
with a finite number $n$ of more or less correlated entries \cite{AllenTildesleyBook,LandauBinderBook}. 
It is supposed here that these series are sampled with a constant time interval 
of length $\delta t$, i.e. $n$ corresponds to a time $t = (n-1) \delta t$.
With $\overline{\Acal_k} \equiv \frac{1}{n} \sum_{k=1}^n \Acal_k$
denoting such a finite average, one samples the $n$-dependent observable \cite{foot_betaV} 
\begin{equation}
\muABn \equiv \beta V \la \overline{(\ahat_k - \overline{\ahat_k}) 
( \bhat_k - \overline{\bhat_k}) } \ra 
\label{eq_muABn_one}
\end{equation}
with $\langle \ldots \rangle$ standing for an additional ensemble average over different
time series of $n$ subsequent data points.
We have $\muABn = 0$ for $n=1$ and $\muABn \to \muAB$ for $n\to \infty$ in an ergodic system.
Interestingly, the detailed time-dependence of $\muABn$ can be obtained from a weighted 
integral of the correlation function $\cABt$ \cite{LandauBinderBook,WXB15}. To see this
we note first the identity
\begin{equation}
\overline{(\ahat_k - \overline{\ahat_k}) ( \bhat_k - \overline{\bhat_k})}
= \frac{1}{2n^2} \sum_{k,l=1}^n (\ahat_k - \ahat_l) (\bhat_k - \bhat_l) 
\label{eq_muABn_two}
\end{equation}
which can be verified by straightforward algebra. (See sect.~2.4 of ref.~\cite{DoiEdwardsBook}.)
Using time translational invariance this implies
\begin{equation}
\muABn = \frac{2}{n^2} \sum_{k=1}^n (n-k) \ g_\mathrm{ab}(\delta t \ k)
\label{eq_muABn_three}
\end{equation}
where the weight $(n-k)$ stems from the finite length of the trajectory used.
We change now to continuous time variables, $k \to s = (k-1) \delta t$, and
replace the discrete sum by the time integral
\begin{equation}
\muABt = \frac{2}{t} \int_0^t \ddiff s \ (1-s/t) \ g_\mathrm{ab}(s).
\label{eq_muABt_gABt}
\end{equation}
Using eq.~(\ref{eq_gABt_cABt}) this yields the general relation
\begin{equation}
1 -\frac{\muABt}{\muAB} =
\frac{2}{t} \int_0^t \ddiff s \ (1-s/t) \ \frac{C_\mathrm{ab}(s)}{C_\mathrm{ab}(0)}  
\label{eq_muABt_cABt}
\end{equation}
which can be used to obtain $\muABt$ if the correlation function $\cABt$ is known.

\paragraph*{Relaxation time $\thetaAB$.}
Interestingly, the time integral in eq.~(\ref{eq_muABt_cABt}) must become
constant in all cases considered below where $\cABt$ vanishes ultimately for $t \to \infty$.
(This applies if a finite shear-barostat is switched on.)
Defining the characteristic relaxation time
\begin{equation}
\thetaAB \equiv \lim_{t\to\infty} \int_0^t \ddiff s \ (1-s/t) \ \frac{C_\mathrm{ab}(s)}{\muAB},
\label{eq_thetaAB_def}
\end{equation}
this leads to
\begin{equation}
1 - \muABt/\muAB \to 2\thetaAB/t \mbox{ for } t \to \infty
\label{eq_muABt_asymp}
\end{equation}
as one expects for the Poisson statistics of uncorrelated events \cite{LandauBinderBook}.
One may thus determine the relaxation time $\thetaAB$ from the large-time
asymptotics of $\muABt$ \cite{foot_theta_f}.

\subsection{Dynamics: Additional model assumptions}
\label{theo_dyna_model}
Let us suppose that the MSD $\gABt$ is diffusive for short times, i.e. $\gABt = \Dab t/2$
for $t \ll \Tab$ with $\Dab$ being a diffusion constant. 
Matching this short time regime with $\gABt = \muAB$ for $t \gg \Tab$ gives the possibility 
to operationally {\em define} $\Tab$ as the crossover time 
\begin{equation}
\Tab \equiv 2 \muAB/\Dab.
\label{eq_TabDab}
\end{equation}
Different short-time dynamics may suggest of course a different operational definition.
Let us further assume an exponentially decaying correlation function 
$\cABt = \muAB \exp(-x)$ with $x = t/\Tab$. Using eq.~(\ref{eq_gABt_cABt}) this
is seen to be consistent with a diffusive short-time regime and eq.~(\ref{eq_TabDab}).
It follows by integration from eq.~(\ref{eq_muABt_cABt}) that
\begin{equation}
1 - \frac{\muABt}{\muAB} = \fDebye(x) \equiv \frac{2}{x^2} [\exp(-x) -1+x]
\label{eq_Debye}
\end{equation}
using the Debye function well-known in polymer science \cite{DoiEdwardsBook}. 
For large reduced times $x \gg 1$ this leads to $1 - \muABt/\muAB \to 2 \Tab/t$,
i.e. by comparison with eq.~(\ref{eq_muABt_asymp}) we have $\Tab = \thetaAB$ for an
exponentially decaying correlation function.

As we shall see in sect.~\ref{simu_dyna_TT}, it may also occur that the correlation function is more or less 
constant between a local (barostat independent time) $\tauA$ up to a (barostat dependent) time $\tstar$, 
i.e. essentially $\cABt \approx c H(\tstar-t)$ with $c$ being a constant and $H(x)$ the Heaviside function. 
It follows from eq.~(\ref{eq_thetaAB_def}) that 
\begin{equation}
1 - \frac{\muABt}{\muAB} \approx \left\{
\begin{array}{ll}
c & \mbox{ for } \tauA \ll t \ll \tstar \\
c  \ \tstar/t & \mbox{ for } \tstar \ll t,
\end{array} 
\right.
\label{eq_intplat}
\end{equation}
i.e. $\thetaAB \approx \tstar \ c/2$ to leading order.

\subsection{Dynamics: Ensemble effects}
\label{theo_dyna_lambda}

\paragraph*{Introduction.}
We have just stated various general relations applying to all ensembles.
These relations do, however, not allow to relate the dynamical correlations 
between different $\lambda$. 
Since some ensembles are more readily computed than others, it would be useful to have a 
transformation relation such as the LPV transform for static fluctuations considered in 
sect.~\ref{theo_stat}.
We emphasize that the correlation functions depend in general on the ensemble,
i.e. the parameter $\lambda$, and on the dynamics of the shear-barostat. 
Since in the end we want to describe the intrinsic relaxation dynamics of the system, it is sufficient to focus
on the limit where the barostat becomes very slow such that it becomes essentially irrelevant for the system 
evolution below an upper time $\tstar$.
We consider the limit where this time $\tstar$ is be much larger than any intrinsic relaxation time of the system. 
Since we want still to consider meaningful thermal averages with respect 
to the chosen ensemble, we need either trajectories of a time interval $\ttraj$ 
much larger than $\tstar$  
to allow the full sampling of the phase space or, equivalently, we need to average
over independent start configurations representing the ensemble.
Under these two assumptions useful transformation relations 
can be formulated by reworking the generalization of the LPV transform
for dynamical correlations replacing the system modulus $\Geq$ by the effective modulus $\Geff$.
Being beyond the scope of this paper, we proceed by postulating the central
scaling relation for the MSD $\gABt$ and argue briefly why this relation is natural.
We discuss then in turn the strain-strain correlations $\cGGt$, the strain-stress 
correlations $\cGTt$ and the stress-stress correlations $\cTTt$ for different $\lambda$. 
Alternative, more direct ways to derive the relations are mentioned.

\paragraph*{Scaling relations.}
We postulate that under the two assumptions made above the MSD $\gABt$ does not depend on the ensemble
\begin{equation}
\gABt \sim \lambda^0 \mbox{ for } t \ll \tstar(\lambda),
\label{eq_MSDscaling}
\end{equation}
i.e. the MSD behaves as a simple mean. This scaling postulate is justified by two facts.
Firstly, the barostat is (by construction) too weak to change for $t \ll \tstar$ the evolution of the system.
Secondly, that the starting points of the trajectory at $t=0$ are distributed according 
to the considered ensemble must become an irrelevant higher order effect
(vanishing rapidly with the system volume $V$), since the MSD probes displacements
with respect to the starting points, not their absolute values. 
The fundamental scaling, eq.~(\ref{eq_MSDscaling}), implies using eq.~(\ref{eq_gABt_cABt}) that
\begin{equation}
\cABt = \muAB(\lambda) - \gABt \mbox{ for } t \ll \tstar(\lambda)
\label{eq_Ctscaling}
\end{equation}
where only the first term on the right hand-side depends on $\lambda$.
This leads us finally to the general transformation relation between correlation
functions 
\begin{equation}
\cABt|_{\lambda} = \cABt|_{\lambda=1} + \muAB|_{\lambda} - \muAB|_{\lambda=1}
\label{eq_Ctlambda}
\end{equation}
where $\lambda=1$ stands for the $\NVgT$-ensemble with $\gamma=\gamext$
The two static contributions on the right hand-side can be further simplified
using results from sect.~\ref{theo_stat}. This demonstrates the
linear relation $\muAB|_{\lambda} - \muAB|_{\lambda=1} \sim 1-\lambda$.

\paragraph*{Strain-strain correlations.}
Since $\gamhat(t) \approx \gamhat(0)$ for $t\ll \tstar$, the MSD $\gGGt$ must 
vanish to leading order. Using eq.~(\ref{eq_Ctscaling}) this implies \cite{foot_tildefunct}
\begin{equation}
\cGGt = \muGG = (1-\lambda)/\Geq \mbox{ for } t \ll \tstar(\lambda).
\label{eq_cGGt_scaling}
\end{equation}
This result is directly obtained by setting $\gamhat(t) = \gamhat(0)$
in the definition of $\cGGt$. We emphasize that it is also consistent with the LPV transform,
eq.~(\ref{eq_LPVgen}), as one verifies by setting $I = \tau$, $X= V\gamma$,
$\Acal=V \gamma(t)$ and $\Bcal = V \gamma(0)$ and using that the strain fluctuation 
term for $\lambda=1$ must vanish. 

\paragraph*{Strain-stress correlations.}
To determine the strain-stress correlation function $\cGTt$ one may use again that
the MSD $\gGTt$ must vanish since $\gamhat(t) \approx \gamhat(0)$ 
for $t \ll \tstar$. Using eq.~(\ref{eq_Ctscaling}) this implies \cite{foot_tildefunct}
\begin{equation}
\cGTt = \muGT = 1-\lambda \mbox{ for } t \ll \tstar(\lambda).
\label{eq_cGTt_scaling}
\end{equation}
This result is also obtained from the LPV transform setting $\Acal=V \gamma(t)$ and 
$\Bcal = \tau(0)$ and using again that $\langle \delta \Ahat \delta \Bhat \rangle|_{\lambda=1} = 0$. 
The postulated scaling eq.~(\ref{eq_MSDscaling}) has thus led again to a reasonable result.

\paragraph*{Stress-stress correlations.}
Interestingly, as one may see from the $\NVgT$-ensemble limit considered in \cite{WXB15,WXBB15}, 
the stress-stress MSD $\gTTt$ is not expected to simply vanish for $t \ll \tstar$ as in the two 
previous cases. 
(The instantaneous stress $\tauhat$ fluctuates even at a fixed strain $\gamhat$.) 
However, eq.~(\ref{eq_Ctlambda}) still holds leading to \cite{foot_tildefunct}
\begin{equation}
\cTTt|_{\lambda} = \cTTt|_{\lambda=1} + (1-\lambda) \Geq
\mbox{ for } t \ll \tstar(\lambda)
\label{eq_cTTt_scaling}
\end{equation}
where the static term on the right hand-side has been simplified using eq.~(\ref{eq_keystat}).
This confirms the key relation eq.~(\ref{eq_keydyna}) announced in the Introduction.
Note that the correlation function $\cTTt|_{\lambda=1}$ in the $\NVgT$-ensemble 
must vanish beyond some local time scale $\tauA$ which does depend on the network considered 
but, of course, not on the shear-barostat.
For a sufficiently slow barostat the correlation function thus becomes constant
\begin{equation}
\cTTt = (1-\lambda) \Geq \mbox{ for } \tauA \ll t \ll \tstar(\lambda)
\label{eq_cTTt_plateau}
\end{equation}
where we have dropped $|_{\lambda}$.
Using eq.~(\ref{eq_Ctscaling}) this leads to the remarkable relation
\begin{equation}
\gTTt = \muA - \Geq \mbox{ for } \tauA \ll t \ll \tstar(\lambda)
\label{eq_gTTt_plateau}
\end{equation}
which must hold for all $\lambda \le 1$. 
We note finally that while for $\lambda<1$
the MSD $\gTTt \to \muTT$ for $t \gg \tstar$, 
$\gTTt = \muA-\Geq$ holds for all times $t \gg \tauA$ for $\lambda=1$
according to stress-fluctuation formula, eq.~(\ref{eq_Gtt}).

\subsection{Dynamics: Macroscopic linear response}
\label{theo_dyna_GtJt}

\paragraph*{Introduction.}
The experimentally important macroscopic linear response measured by the creep 
compliance $J(t)$ and the shear relaxation modulus $G(t)$ 
\cite{DoiEdwardsBook,WittenPincusBook,RubinsteinBook}
may be obtained conveniently in an equilibrium simulation at a given $\lambda$
using some of the correlation functions discussed above.
Please note that being material properties of the given state point 
(experimentally obtained using a simple average, not a fluctuation) 
both response functions $J(t)$ and $G(t)$ do, of course, not depend on $\lambda$.

\paragraph*{Creep compliance.}
Let us first consider the creep compliance 
$J(t) \equiv \la \delta \gamhat(t) \ra/\delta \tauext$ for $t \ge 0$.
It is assumed that for $t < 0$ the system is at thermal equilibrium
and the internal mean stress $\tau$ equals the applied external stress $\tauext$
of the $\NVtT$-ensemble. After imposing at $t = 0$ a small increment $\delta \tauext$, 
the creep compliance $J(t)$ measures the ensuing average strain increment 
$\la \delta \gamhat(t) \ra$. We note {\em en passant} that the average internal shear 
stress $\la \tauhat(t) \ra$ does neither immediately reach the new equilibrium
value $\tauext + \delta \tauext$ but shows a similar time dependence as the strain.
Reworking the arguments put forward by Doi and Edwards, see eq.~(3.67) of ref.~\cite{DoiEdwardsBook},
it is seen that
\begin{equation}
J(t) = \gGGt|_{\lambda=0} \mbox{ for } |\delta \tauext| \ll 1,
\label{eq_Jt}
\end{equation}
i.e. the creep compliance is most readily computed using the strain-strain MSD $\gGGt$
in the $\NVtT$-ensemble.
As described in sect.~\ref{algo_finiteT}, we shall change the shear-strain $\gamhat(t)$
using a MC shear-barostat which corresponds to a perfectly viscous dynamics.
One thus expects $\cGGt = \muGG \exp(-t/\Tgg)$ for the strain-strain correlations.
Using $\muGG=1/\Geq$ for $\lambda=0$ and eq.~(\ref{eq_gABt_cABt}) this suggest 
\begin{equation}
J(t) = \frac{1}{\Geq} \ [1-\exp(-t/\Tgg)]
\label{eq_KelvinVoigt}
\end{equation}
in agreement with the Kelvin-Voigt model \cite{RubinsteinBook} 
representing a purely viscous damper and a purely elastic spring connected in parallel.

\paragraph*{Shear relaxation modulus.}
The shear relaxation modulus $G(t) \equiv \la \delta \tauhat(t) \ra/\delta \gamma$
may be obtained from the stress increment $\la \delta \tauhat(t) \ra$ for $t > 0$
after a small step strain with $|\delta \gamma| \ll 1$ has been imposed at time $t=0$.
It is well known that the components of the Fourier transformed relaxation modulus $G(t)$,
the storage modulus $\Gstor$ and the loss modulus $\Gloss$, are directly measurable in 
an oscillatory shear strain experiment \cite{RubinsteinBook,WXBB15}.
As seen, e.g., by eq.~(32) in ref.~\cite{WXBB15}, it can be demonstrated 
by integration by parts that
\begin{equation}
G(t) = \cTTt|_{\lambda=0} \mbox{ for } t \ll \tstar(\lambda=0),
\label{eq_Gt_NVtT}
\end{equation}
i.e. the barostat should be irrelevant on the time scales considered.
Using the transformation relation between different ensembles, eq.~(\ref{eq_cTTt_scaling}),
the relaxation modulus may equivalently be obtained for other $\lambda$
according to
\begin{equation}
G(t) = \cTTt|_{\lambda} + \lambda \Geq \mbox{ for } t \ll \tstar(\lambda).
\label{eq_Gt_lambda}
\end{equation}
Since for large times $G(t) \to \Geq$, this implies $\cTTt|_{\lambda} \to (1-\lambda) \Geq$
consistently with eq.~(\ref{eq_cTTt_plateau}).
For the specific case $\lambda=1$ eq.~(\ref{eq_Gt_lambda}) yields 
\begin{equation}
G(t) = \cTTt|_{\lambda=1} + \Geq
\label{eq_Gt_NVgT}
\end{equation} 
which holds for all times $t$ since the barostat becomes irrelevant for $\lambda \to 1$.
Note that eq.~(\ref{eq_Gt_NVgT}) may also be derived directly (without using
the transformation relation between different ensembles) using 
Boltzmann's superposition principle for an arbitrary strain history
and the standard fluctuation-dissipation theorem for the after-effect function 
\cite{HansenBook} as shown elsewhere \cite{WXB15,WXBB15}.
Two immediate consequences of eq.~(\ref{eq_Gt_NVgT}) are 
{\em (i)} 
that $G(t)$ only becomes equivalent to $\cTTt|_{\lambda=1}$ for $t > 0$ in the liquid limit 
where (trivially) $\Geq=0$ and {\em (ii)} that the shear modulus $\Geq$ is only 
probed by $G(t)$ on time scales $t \gg \tauA$ where $\cTTt|_{\lambda=1}$ must vanish.  
In principle, it is thus {\em impossible} to obtain the static shear modulus $\Geq$ 
of an elastic body only from $\cTTt|_{\lambda=1}$ as often incorrectly assumed 
\cite{Klix12,Szamel15}.

\section{Algorithmic details}
\label{sec_algo}

\subsection{Model Hamiltonian}
\label{algo_model}

To illustrate our key relations we present in sect.~\ref{sec_simu} numerical data obtained 
using a periodic two-dimensional ($d=2$) network of $N_l$ harmonic springs connecting
$N$ vertices.
The model Hamiltonian is given by the sum $\Hhat = \Hidhat + \Hexhat$ of 
a kinetic energy contribution 
\begin{equation}
\Hidhat = \frac{m}{2} \sum_{i=1}^N \vvec_i^2,
\label{eq_Eid}
\end{equation}
with $\vvec_i$ being the velocity of vertex $i$ and $m$ its (assumed) monodisperse mass, 
and an excess potential 
\begin{equation}
\Hexhat = \sum_{l=1}^{N_l} \ul(\rl) \mbox{ with } \ul(r) = \frac{1}{2} K_l \left(r - R_l\right)^2
\label{eq_Enet}
\end{equation}
where $K_l$ denotes the spring constant, $R_l$ the reference length and $\rl = |\rvec_i - \rvec_j|$ 
the length of spring $l$. The sum runs over all springs $l$ connecting pairs of beads $i$ and $j$ 
with $i < j$ at positions $\rvec_i$ and $\rvec_j$. 
The vertex mass $m$ and Boltzmann's constant $\kB$ are set to unity and
Lennard-Jones (LJ) units \cite{AllenTildesleyBook} are assumed.

\subsection{Canonical affine shear transformations}
\label{algo_affine}

While the box volume $V$ is kept constant throughout this work, we shall frequently
change the shape of the simulation box. As sketched in panel (a) of fig.~\ref{fig_sketch}, 
we perform plane shear transformations of the instantaneous shear strain 
$\gamhat \to \gamhat + \delta \gamma$ with an essentially infinitesimal
strain increment $\delta \gamma$.
We assume that not only the box shape is changed but that the particle positions $\rvec$
(using the principal box convention \cite{AllenTildesleyBook}) follow the imposed macroscopic 
constraint in an {\em affine} manner according to 
\begin{equation}
\rx \to \rx + \delta \gamma \ \ry \ \mbox{ for } |\delta \gamma| \ll 1
\label{eq_r_cantrans}
\end{equation}
with all other coordinates remaining unchanged.
Albeit not strictly necessary for the demonstration of our key relations, 
we assume, moreover, that this shear transformation is also {\em canonical} 
\cite{Goldstein,FrenkelSmitBook}. This implies that the $x$-component of the velocity
must transform as
\cite{WXBB15}
\begin{equation}
\vx \to \vx - \delta \gamma \ \vy \ \mbox{ with } |\delta \gamma| \ll 1.
\label{eq_p_cantrans}
\end{equation}
We emphasize the negative sign in eq.~(\ref{eq_p_cantrans}) 
which assures that Liouville's theorem is obeyed \cite{Goldstein,WXBB15}.

\subsection{Shear stress and affine shear-elasticity}
\label{algo_tau_muA}
Let $\Hhat(\delta \gamma) = \Hidhat(\delta \gamma) + \Hexhat(\delta \gamma)$ 
denote the system Hamiltonian of a configuration originally at $\gamhat$ 
strained using a canonical affine transformation to $\gamhat + \delta \gamma$ 
compactly written as a function of the strain increment $\delta \gamma$.
The instantaneous shear stress $\tauhat$ and the instantaneous affine shear-elasticity $\muAhat$
may be defined as the expansion coefficients associated to the energy change
\begin{equation}
\delta \Hhat(\delta \gamma)/V = \tauhat \delta \gamma + \muAhat \delta \gamma^2/2 
\mbox{ for } |\delta \gamma| \ll 1
\label{eq_Haffine}
\end{equation}
with $\gamhat$ being the reference, i.e. \cite{foot_prime}
\begin{eqnarray}
\tauhat & \equiv & \Hhat^{\prime}(\delta \gamma)/V|_{\gamhat} \label{eq_tauhatdef} \mbox{ and } \\
\muAhat & \equiv & \Hhat^{\prime\prime}(\delta \gamma)/V|_{\gamhat} \label{eq_muAhatdef}.
\end{eqnarray}
The derivatives $\Hhat^{\prime}(\delta \gamma)$ and $\Hhat^{\prime\prime}(\delta \gamma)$ 
with respect to $\delta \gamma$ may be computed as shown in sect.~2.1 of \cite{WXBB15}.
Similar relations apply for the corresponding contributions
$\tauidhat$ and $\tauexhat$ to  $\tauhat =\tauidhat + \tauexhat$ and 
for the contributions $\muAidhat$ and $\muAexhat$ to $\muAhat = \muAidhat + \muAexhat$. 
Using eq.~(\ref{eq_Eid}) this implies \cite{WXBB15}
\begin{eqnarray}
\tauidhat & = & - \frac{1}{V} \sum_{i=1}^N m_i \vix \viy \label{eq_tauidhat} \mbox{ and } \\
\muAidhat & = & \frac{1}{V} \sum_{i=1}^N m_i \viy^2 \label{eq_muAidhat} 
\end{eqnarray}
for the ideal contributions to the shear stress and the affine shear-elasticity.
Note that the minus sign for the shear stress is due to the minus sign in eq.~(\ref{eq_p_cantrans})
required for a canonical transformation.
For the excess contributions one obtains \cite{WXBB15}
\begin{eqnarray}
\tauexhat & = & \frac{1}{V} \sum_l \rl u^{\prime}(\rl) \ \nlx \nly   \label{eq_tauexhat} \ \mbox{ and } \\
\muAexhat & = & \frac{1}{V} \sum_l  \left( \rl^2 u^{\prime\prime}(\rl)
- \rl u^{\prime}(\rl) \right) \nlx^2 \nly^2 \nonumber \\
& + & \frac{1}{V} \sum_l \rl u^{\prime}(\rl) \ \nly^2  \label{eq_muAexhat}
\end{eqnarray}
with $\nvecl = \rvecl/\rl$ being the normalized distance vector $\rvec= \rvec_j - \rvec_i$
between the particles $i$ and $j$.
Interestingly, eq.~(\ref{eq_tauexhat}) is strictly identical to the 
corresponding off-diagonal term of the Kirkwood stress tensor  \cite{AllenTildesleyBook}.
The last term in eq.~(\ref{eq_muAexhat}) automatically takes into account the finite 
normal pressure of the system.

\begin{figure}[t]
\centerline{
\resizebox{1.0\columnwidth}{!}{\includegraphics*{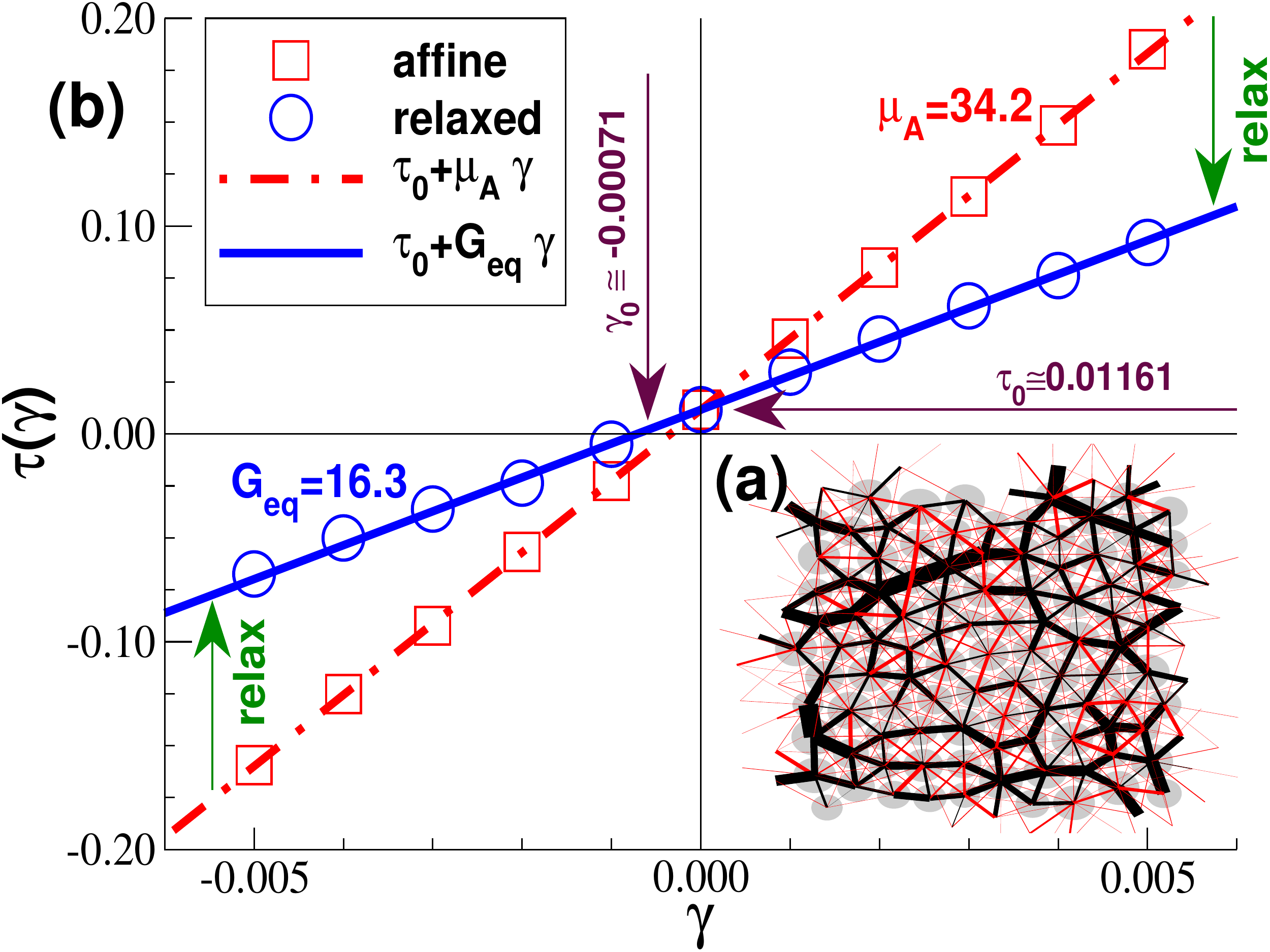}}
}
\caption{Groundstate of elastic network model considered in this work assuming
eq.~(\ref{eq_Enet}):
{\bf (a)}
Snapshot of a small subvolume of linear length $10$ containing about $100$ vertices.
The lines represent the quenched forces of the athermal ($T=0$) reference configuration.
{\bf (b)}
Shear stress $\tau(\gamma)$ assuming an affine strain (squares)
and after energy minimization (circles). In agreement with ref.~\cite{WXBB15}
we find in the first case $\tau(\gamma) = \tauzero + \muA \gamma$ with $\muA = 34.2$ (dashed line)
and in the second case $\tau(\gamma) = \tauzero + \Geq \gamma$ with $\Geq = 16.3$ (bold solid line).
The stress does not vanish at $\gamma=0$ where $\tau = \tauzero =0.01161$
(horizontal arrow) but at $\gamma = \gamzero = -0.00071$ (vertical arrow).
\label{fig_algo}
}
\end{figure}

\subsection{Groundstate characterization}
\label{algo_refer}

\paragraph*{Specific network.}
As explained elsewhere \cite{WXP13,WXBB15}, the specific elastic network used in this work
has been constructed using the dynamical matrix of a quenched polydisperse LJ bead glass, 
i.e. at low temperatures our network has exactly the same mechanical and vibrational 
properties as the original discrete particle system.
Prior to forming the network the  bead system was cooled down to $T=0$ using a constant
quenching rate and imposing a normal pressure $P=2$.
The original LJ beads are represented in the snapshot shown in fig.~\ref{fig_algo}
by grey polydisperse circles, the permanent spring network created from the quenched
bead system by lines between vertices.
The dark (black) lines indicate repulsive forces between the vertices,
while the light (red) lines represent tensile forces.
The line width is proportional to the tension (repulsion).
Note that the force network is strongly inhomogeneous with zones
of weak attractive links embedded within a strong repulsive skeleton
as already discussed in refs.~\cite{WTBL02,TWLB02}.
Only a small subvolume of the network is represented.
The total periodic box of linear length $L \approx 102.3$
contains $N=10^4$ vertices and $N_l=9956$ springs.
The monomer density $\rho$ is close to unity.

\paragraph*{Finite shear stress $\tauzero$.}
Due to the construction of the network the total force acting on each vertex 
of the reference network must vanish at $T=0$. As seen in the snapshot, this does
not imply that the repulsive and/or tensile forces transmitted along the springs 
must also vanish.
Due to the periodic boundary conditions and the constant-strain constraint ($\gamma=0$) 
the shear stress $\tau$ does not necessarily vanish. For a pair potential such as 
eq.~(\ref{eq_Enet}) the relevant excess contribution $\tauexhat$ of the shear stress 
is readily computed using the Kirkwook expression, eq.~(\ref{eq_tauexhat}).
As shown in the main panel of fig.~\ref{fig_algo} (horizontal arrow), it turns out that 
for the specific network we use throughout this work we have a finite shear stress 
$\tauzero \equiv \tau(\gamma=0) = 0.01161$.

\paragraph*{Affine shear-elasticity $\muA$.}
Let us consider a small affine shear strain according to eq.~(\ref{eq_r_cantrans}).
As show in panel (b) of fig.~\ref{fig_algo}, one may now compute using eq.~(\ref{eq_tauexhat}) 
the shear stress $\tau(\gamma)$ for different $\gamma$ (squares). As one expects from 
the definition of the affine shear-elasticity coefficient $\muA$, eq.~(\ref{eq_Haffine}),
this yields the linear relation 
$\tau(\gamma) = \tauzero + \muA \gamma$ with $\muA \approx 34.2$ 
as indicated by the dashed line. The same coefficient $\muA$ is also obtained
directly from the unstrained configuration using eq.~(\ref{eq_muAexhat}).

\paragraph*{Equilibrium shear modulus $\Geq$.}
The forces $\fvec_i$ acting on the vertices $i$ of an {\em affinely} strained network
do not vanish in general and the system is normally {\em not} at mechanical equilibrium. 
As described elsewhere \cite{TWLB02,WXBB15}, 
we relax these forces by first applying a steepest descend algorithm, 
i.e. by imposing displacements proportional to the force,
and then by means of the conjugate gradient method \cite{ThijssenBook}.
The ensuing non-affine displacements of the vertices decrease the energy \cite{WXBB15} and
the magnitude of the shear stress $\tau(\gamma)$ as may be seen from the large circles indicated 
in fig.~\ref{fig_algo}. As indicated by the bold solid line 
these final stresses $\tau(\gamma)$ scale again linearly as
\begin{equation}
\tau(\gamma) = \tauzero + \Geq \gamma = \Geq (\gamma - \gamzero)
\label{eq_gamzero_def}
\end{equation}
with $\Geq \approx 16.3$ and $\gamzero = - \tauzero/\Geq \approx -0.00072$. 
The affine coefficient $\muA$ has thus been replaced by the much smaller equilibrium shear modulus
$\Geq$ of the groundstate network. Note that shear stress vanishes at a strain 
$\gamma = \gamzero$ as indicated by the vertical arrow. 
If the strain $\gamma$ is allowed to change freely (e.g., using a steepest descend scheme)
without any external force $\tauext$ applied, the system relaxes to $\gamma=\gamzero$.

\subsection{Finite temperature simulations}
\label{algo_finiteT}

\paragraph*{Molecular dynamics scheme.}
As discussed in sect.~\ref{sec_simu}, this network is investigated numerically basically
by means of a molecular dynamics (MD) simulation 
\cite{AllenTildesleyBook,FrenkelSmitBook} at constant particle number $N=10^4$, 
box volume $V \approx 102.3^2$ and a small, but finite mean temperature $T=0.001$.
Newton's equations are integrated using a velocity-Verlet algorithm with a tiny 
time step $\dtMD=10^{-4}$. The temperature $T$ is fixed using a Langevin thermostat 
with a relatively large friction constant $\zeta =1$.
This was done to suppress artificial long-range correlations in the two-dimensional periodic simulation box.

\paragraph*{Thermal averages.}
Using the measured instantaneous shear stress $\tauhat$ and affine shear-elasticity $\muAhat$
we compute, e.g., the thermal averages $\tau \equiv \la \tauhat \ra$, 
$\muTT \equiv \beta V \la \delta \tauhat^2 \ra$ and $\muA \equiv \la \muAhat \ra$.
It can be shown for the respective ideal contributions that \cite{WXBB15}
\begin{equation}
\tauid = \muTTid = \muAid = \Pid = T \ \rho 
\label{eq_thermal_ideal}
\end{equation}
with $\Pid$ being the ideal normal pressure contribution. 
This holds irrespective of the considered $\lambda$-ensemble.
The ideal contributions are thus negligible.
Due to the equivalence of the different axes it is also seen for the excess contributions 
that \cite{WXBB15}
\begin{eqnarray}
\muAex & = & \muB - \Pex   \mbox{ with } \nonumber \\
\muB & = & \frac{1}{V} \sum_l 
\la \left( \rl^2 u^{\prime\prime}(\rl) - \rl u^{\prime}(\rl) \right) \nlx^2 \nly^2 \ra
\label{eq_muA}
\end{eqnarray}
being the Born-Lam\'e coefficient \cite{BornHuang,Barrat88,Lutsko89,WTBL02,TWLB02,Barrat13}
and $\Pex$ the excess part of the normal pressure $P=\Pid+\Pex$.

\begin{table}[t]
\begin{tabular}{|c||c|c|c||c|c|c|c|c|}
\hline \hline
$\Gext$   &
$\lambda$ & 
$\muGG$   & 
$\muTT$   & 
$A$  &
$\eta$  &
$\Tgg$  &
$\Tgt$  &
$\thetaTT$ 
\\ \hline
-10   & -1.59 & 0.159 & 60.1 & 0.997 & 0.0081  & 9.2  & 4.3 & 13 \\
-6    & -0.58 & 0.097 & 43.7 & 0.996 & 0.0103  & 5.6  & 2.7 & 6\\
-3    & -0.23 & 0.075 & 37.9 & 0.996 & 0.0117  & 4.3  & 2.1 & 4\\
0     & 0     & 0.061 & 34.2 & 0.996 & 0.0130  & 3.5  & 1.7 & 2 \\
16.3  & 0.50  & 0.031 & 26.1 & 0.994 & 0.0184  & 1.8  & 0.85 & 1 \\
100   & 0.86  & 0.009 & 20.1 & 0.993 & 0.0348  & 0.5  & 0.24 & 0.3\\
1000  & 0.98  & 0.001 & 18.1 & 0.979 & 0.1030  & 0.06 & 0.03 & 0.3 \\
10000 & 0.998 & -     & 17.9 & 0.935 & 0.3233  & 0.007& 0.003 & 0.4 \\
\hline
\end{tabular}
\vspace*{0.5cm}
\caption[]{Summary of some properties as a function of the external spring constant $\Gext$:
$\lambda=\Gext/(\Geq+\Gext)$ with $\Geq=16.3$, 
strain-strain fluctuation $\muGG$ (fig.~\ref{fig_dgam}), 
stress-stress fluctuation $\muTT$ (fig.~\ref{fig_muF_L}),
acceptance rate $A$ of MC shear-barostat (fig.~\ref{fig_Aeta}),
$\eta(\lambda,\kappa)$-parameter defined by eq.~(\ref{eq_eta_def}),
strain-strain crossover time $\Tgg \approx \thetaGG$ (fig.~\ref{fig_gGGt}),
strain-stress crossover time $\Tgt \approx \thetaGT$ (fig.~\ref{fig_gGTt}),
stress-stress relaxation time $\thetaTT$ (fig.~\ref{fig_thetaTT}).
The dynamical properties (columns 5-9) are only given for $\kappa=10^{-2}$.
The time scales $\Tgg$, $\Tgt$ and $\thetaTT$ should be compared to the intrinsic relaxation time
$\tauA \approx 0.13$ of the stress-stress correlations for $\kappa \to 0$
(fig.~\ref{fig_gTTt}).
\label{tab}}
\end{table}

\paragraph*{Monte Carlo shear-barostat.}
The MD algorithm for the particles is coupled for $\lambda < 1$
with a Monte Carlo (MC) scheme \cite{AllenTildesleyBook,LandauBinderBook} 
attempting every MD time step a canonical affine strain 
increment $\gamhat \rightarrow \gamhat + \delta \gamma$ (sect.~\ref{algo_affine}).
First, a strain increment $\delta \gamma$ is randomly chosen from a uniformly
distributed interval $[-\gammamax,\gammamax]$.
In order to determine the Metropolis weight \cite{LandauBinderBook} 
we compute next the energy change 
$\delta \Hhat = \Hhat(\gamhat + \delta \gamma) - \Hhat(\gamhat)$ of the network 
which comprises both an excess contribution due to eq.~(\ref{eq_r_cantrans}) and 
an ideal contribution due to eq.~(\ref{eq_p_cantrans}). 
Since our system is subjected to an external field $\Uext(\gamhat)$, eq.~(\ref{eq_Uext}), 
this gives rise to an additional contribution 
$\delta \Uext = \Uext(\gamhat+\delta \gamma) - \Uext(\gamhat)$.
The suggested strain move $\delta \gamma$ is accepted if
\begin{equation}
\xi \le \exp[-\beta (\delta \Hhat + \delta \Uext)]
\label{eq_Metropol}
\end{equation}
with $\xi$ being a uniformly distributed random variable with $0 \le \xi < 1$
\cite{LandauBinderBook}.

\begin{figure}[t]
\centerline{\resizebox{1.0\columnwidth}{!}{\includegraphics*{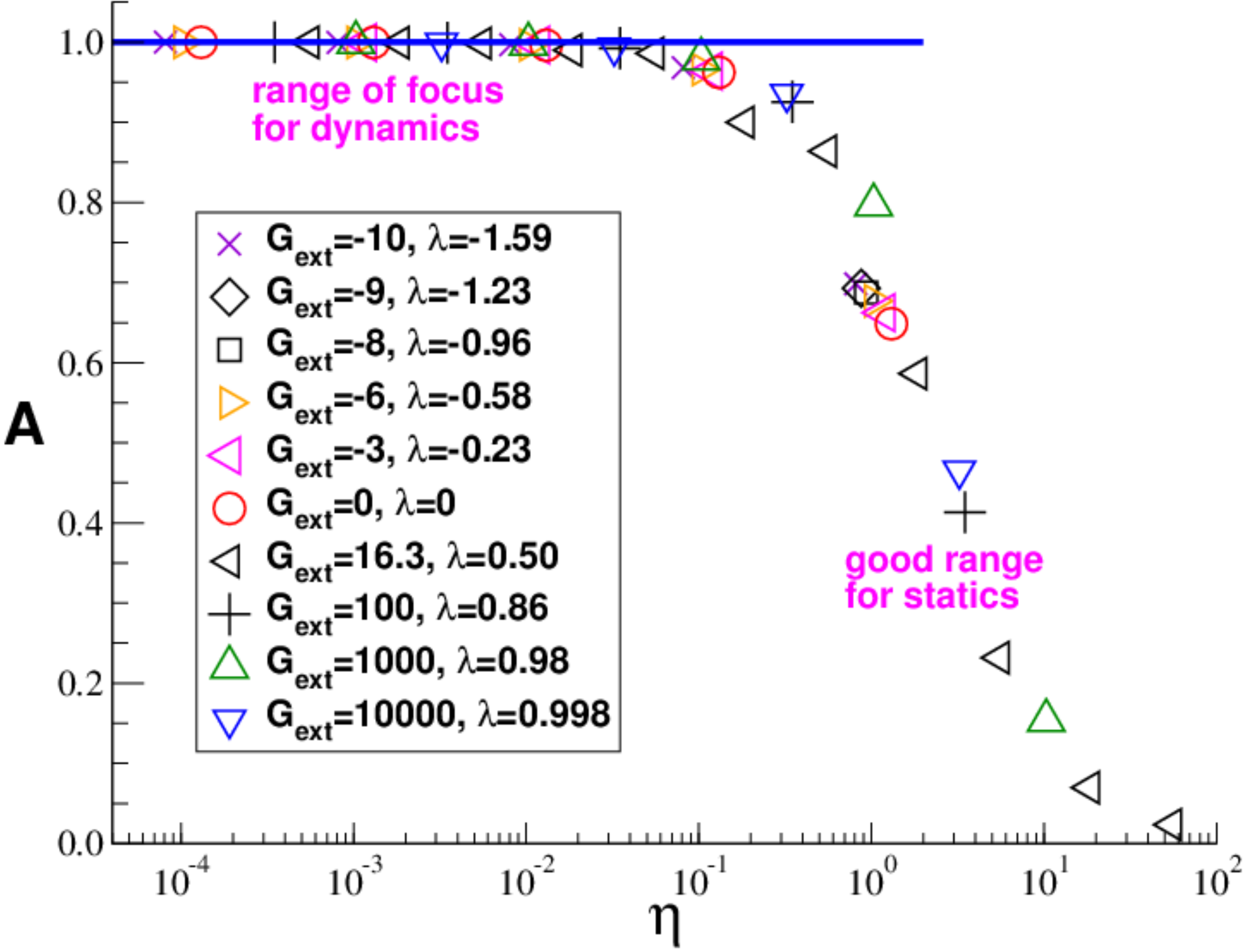}}}
\caption{Acceptance rate $A$ of the Metropolis MC shear-barostat, eq.~(\ref{eq_Metropol}),
for a large range of $\lambda$ and $\kappa$ as a function of the
scaling variable $\eta(\lambda,\kappa)$ as defined by eq.~(\ref{eq_eta_def}).
\label{fig_Aeta}
}
\end{figure}

\paragraph*{Operational parameters $\lambda$ and $\kappa$.}
%
We assume for the external field $\Uext(\gamhat)$ that $\tauext \equiv 0$ and 
$\gamext \equiv \gamzero$, i.e. the average shear stress $\tau$ is imposed to vanish 
for all ensembles studied and all ensembles compare the same thermodynamic state point.
(This was explicitly checked.)
The only remaining operational parameter from the static point of view is the
external spring constant $\Gext$ or, equivalently, $\lambda \equiv \Gext / (\Geq + \Gext)$ 
as sketched in panel (c) of fig.~\ref{fig_sketch}. 
As seen in table~\ref{tab} or fig.~\ref{fig_Aeta}, we vary $\Gext$ from $-10$, 
i.e. $\lambda \approx -1.59$,
over $\Gext=\lambda=0$ ($\NVtT$-ensemble) up to $\Gext =10000$, i.e. $\lambda \approx 0.998$.
The latter case is essentially equivalent to the standard $\NVgT$-ensemble,
i.e. the strain fluctuations become irrelevant for most properties.
The second operational parameter of this study is the maximum attempted strain displacement 
$\gammamax$ which determines the impact of the shear-barostat on the relaxation dynamics.
Since the Metropolis MC move for the strain is performed every MD time step of length $\dtMD$,
it is convenient to use instead of $\gammamax$ the maximum strain increment {\em rate} 
$\kappa \equiv \gammamax/\dtMD$. (Since all simulations are performed with
the same $\dtMD$, $\gammamax$ and $\kappa$ are strictly equivalent.)
We compare below the dynamical strain and stress correlations for the five rates 
$\kappa= 1, 10^{-1}, 10^{-2}, 10^{-3}$ and $10^{-4}$ for a broad range of $\lambda$.
For $\lambda=0.5$ we have computed in addition the values $\kappa = 3, 0.3, 0.03, 0.003$ and $0.0003$.
Some properties for $\kappa=10^{-2}$ are summarized in the columns 5-9 of table~\ref{tab}. 
The simulations become increasingly time consuming with decreasing $\kappa$
and data obtained for $\kappa=10^{-4}$ have to be taken with care.

\paragraph*{Free strain diffusion limit.}
The impact of $\kappa$ can be best judged from the simple limit where neither the system nor 
the external field restricts the barostat. 
This is the case for sufficiently small $\kappa$ depending on $\lambda$. 
In this limit one expects the free diffusion of the strain 
$\gamhat(t)$, i.e. $\gGGt = \Dgg t/2$ for $t \ll \Tgg$.
Since the (attempted and accepted) strain increment $\delta \gamma$ is uniformly 
distributed in $[-\gammamax,\gammamax]$, we have a mean-square strain step
$\langle \delta \gamma^2 \rangle = \gammamax^2/3$ every $\dtMD$.
This implies a diffusion constant
\begin{equation}
\Dgg = \frac{\beta V}{3} \ \frac{\gammamax^2}{\dtMD} \sim \kappa^2
\label{eq_Dgg_FDL}
\end{equation}
in the free-diffusion limit. Using eq.~(\ref{eq_TabDab}) and eq.~(\ref{eq_dgam_L}) 
this yields in turn the corresponding crossover time 
\begin{equation}
\Tgg = 6 \frac{\muGG}{\beta V} \frac{\dtMD}{\gammamax^2} = 6 \ \dtMD/\eta^2
\label{eq_Tgg_FDL}
\end{equation}
where we have defined the dimensionless variable 
\begin{equation}
\eta(\lambda,\kappa) \equiv 
\sqrt{\beta V \gammamax^2/\muGG} = \frac{\kappa \ \dtMD}{\sqrt{\langle \delta \gamhat^2 \rangle}}.
\label{eq_eta_def}
\end{equation}
Since $\Geq$ and $\dtMD$ are kept constant in all our simulations, 
the parameter $\eta$ determines the dynamical regime for a system 
of operational parameters $\lambda$ and $\kappa$.
As seen in fig.~\ref{fig_Aeta} using $\eta$ as a scaling variable the acceptance
rate $A$ of the MC shear-barostat of a broad range of $\lambda$ and $\kappa$ can be brought to collapse.
For $1 \ll \eta \ll 10$ the MC barostat is most efficient for static properties.
For dynamical properties we shall focus below on $\kappa$-values where $\eta \ll 0.1$ 
and thus $A \approx 1$ as emphasized by the solid horizontal line.

\section{Computational results}
\label{sec_simu}

\subsection{Static properties}
\label{simu_stat}

\begin{figure}[t]
\centerline{\resizebox{1.0\columnwidth}{!}{\includegraphics*{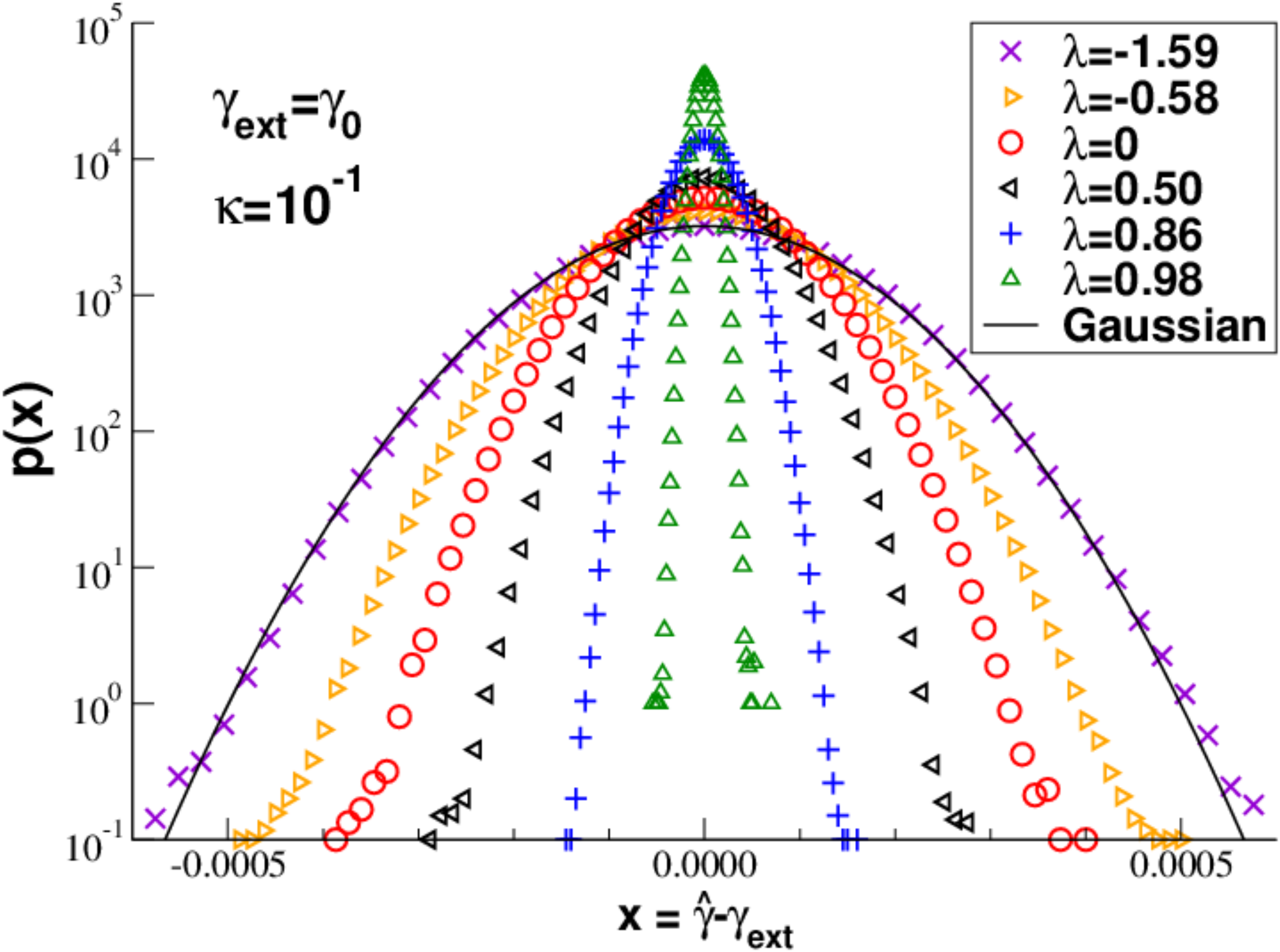}}}
\caption{Normalized histogram $p(x)$ of $x = \gamhat-\gamext$ for different $\lambda$.
The line indicates for $\lambda=-1.59$ ($\Gext=-10$) a Gaussian with 
$\la \delta \gamhat^2 \ra = 1/\beta V \Geff \approx 0.00012^2$.
\label{fig_histoL}
}
\end{figure}

\begin{figure}[t]
\centerline{\resizebox{1.0\columnwidth}{!}{\includegraphics*{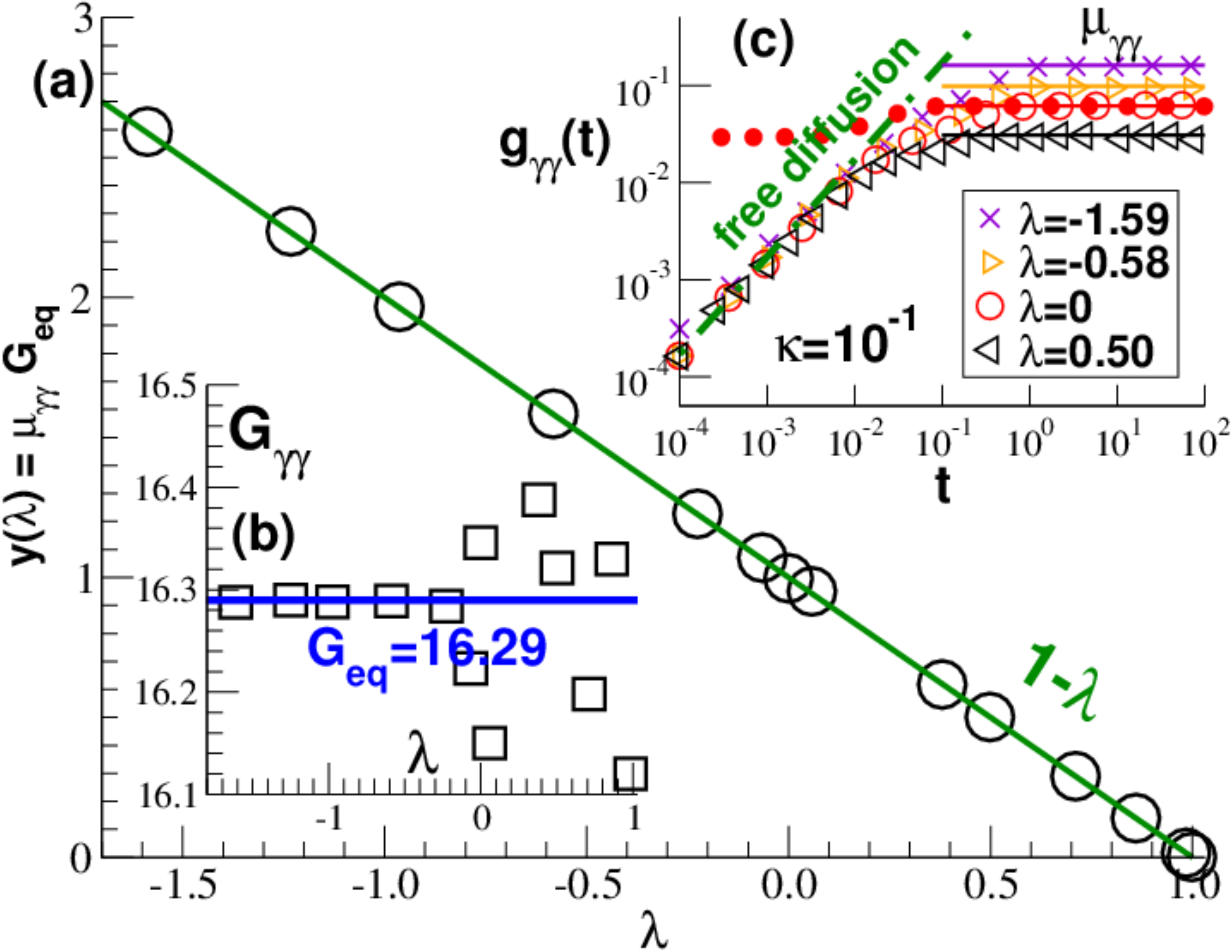}}}
\caption{Second moment of strain fluctuations:
{\bf (a)} 
Dimensionless $y = \muGG \Geq$ with $\Geq\equiv 16.3$ as a function of $\lambda$ showing
the linear decay (bold line) expected from eq.~(\ref{eq_dgam_L}).
{\bf (b)} Shear modulus $\Ggg$, eq.~(\ref{eq_Ggggen}), confirming $\Geq \approx 16.3$ (bold line).
{\bf (c)}
MSD $\gGGt$ {\em vs.} time $t$ for several $\lambda$ and $\kappa=10^{-1}$.
The dash-dotted line indicates the free-diffusion limit, eq.~(\ref{eq_Dgg_FDL}), for short times,
the horizontal lines the long-time limit $\muGG$ for each $\lambda$.
Also indicated is for $\lambda=0$ and $\kappa=1$ (filled circles)
an example with $\eta \gg 0.1$ where the initial diffusive regime is suppressed. 
\label{fig_dgam}
}
\end{figure}

\paragraph*{Strain-strain fluctuations.}
The equilibrium fluctuations of the strain $\gamhat$ at a finite temperature $T=0.001$ 
for different values of $\Gext$ or, equivalently, $\lambda=\Gext/(\Geq+\Gext)$ with $\Geq \equiv 16.3$ 
are presented in fig.~\ref{fig_histoL} and fig.~\ref{fig_dgam}. 
The given data refer to a maximum strain rate $\kappa = 10^{-1}$ for the
Metropolis MC step $\gamhat \to \gamhat + \delta \gamma$ 
attempted after each MD time step of $\dtMD = 10^{-4}$  (sect.~\ref{algo_finiteT}). 
That this leads indeed to the desired Gaussian distributions can be seen 
from fig.~\ref{fig_histoL} where the normalized histogram $p(x)$ is plotted as a function 
of $x = \gamhat-\gamext$ with $\gamext=\gamzero$.
As one expects, $p(x) \to \delta(x)$ for $\lambda \to 1$. 
The rescaled second moment $\muGG \equiv \beta V \langle \delta \gamhat^2 \rangle$ 
of the distribution is further analyzed in fig.~\ref{fig_dgam}. As one may see
from panel (a), $y(\lambda) \equiv \muGG \Geq$
decreases linearly with $\lambda$ (bold solid line) and vanishes for $\lambda =1$ as it should
according to eq.~(\ref{eq_muGG_lambda}).
In turn this implies that one may determine $\Geq$ from 
\begin{equation}
\Ggg \equiv 1/\muGG - \Gext.
\label{eq_Ggggen}
\end{equation}
As can be seen from panel (b) of fig.~\ref{fig_dgam}, using this relation one confirms that the 
groundstate value $\Geq \approx 16.3$ remains unchanged at a finite temperature $T \ll 1$.
The accurate determination of $\Ggg$ becomes of course more delicate with increasing $\lambda$
since the MC acceptance rate becomes eventually too small. 
The data quality in this limit can be readily improved (not shown) by either 
increasing the sampling time $\ttraj$ or by decreasing $\kappa \sim \eta$.
(The sampling becomes again inefficient if $\eta$ gets too small.)
\begin{figure}[t]
\centerline{\resizebox{1.0\columnwidth}{!}{\includegraphics*{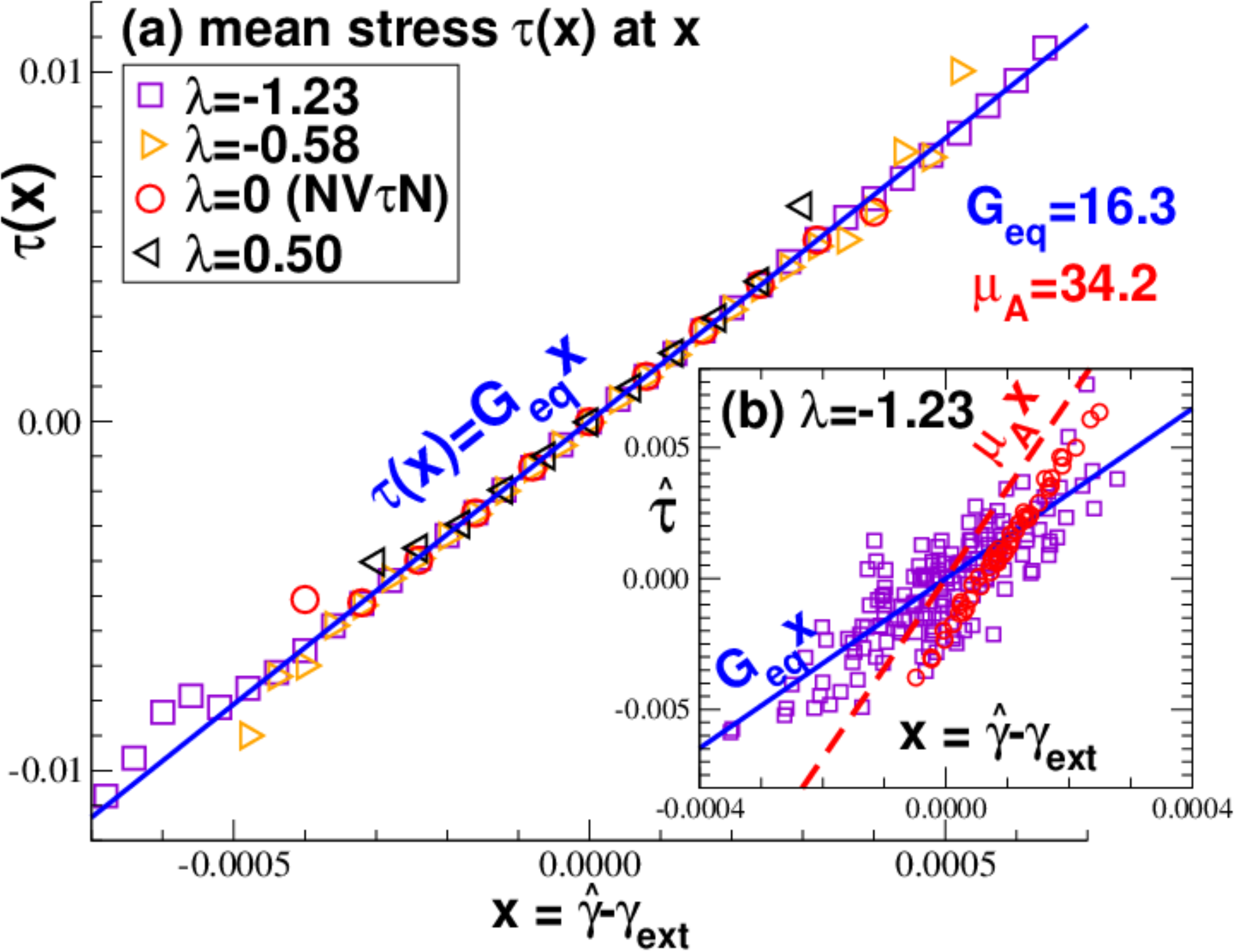}}}
\caption{Strain-stress correlations:
{\bf (a)}
Pre-averaged shear stress $\tau(x)$ as a function of $x=\gamhat-\gamext$.
All data collapse on the linear slope with $\Geq=16.3$ (bold solid line).
{\bf (b)}
Scatter plot of 100 data points $(\gamhat,\tauhat)$ for $\lambda=-1.23$ ($\Gext=-9$) 
with circles corresponding to a tiny time interval $\delta t = 10^{-4}$ (dashed line)
and squares to a large $\delta t = 10^{+2}$ (bold solid line) between subsequent configurations.
\label{fig_tau_gam}
}
\end{figure}

\begin{figure}[t]
\centerline{\resizebox{1.0\columnwidth}{!}{\includegraphics*{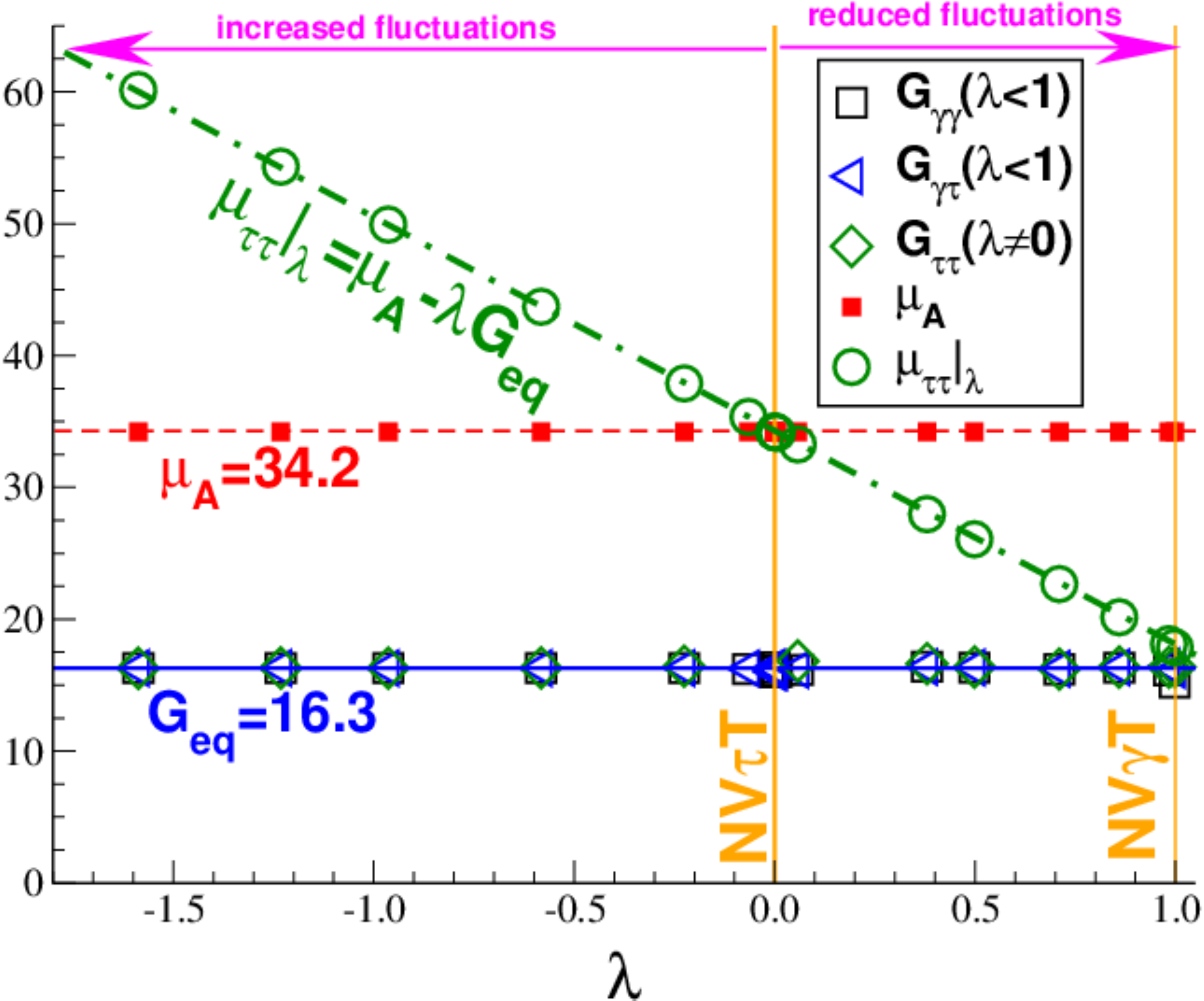}}}
\caption{Determination of shear modulus $\Geq \approx 16.3$ (bold solid line)
using $\Ggg$ for $\lambda < 1$, $\Ggt$ for $\lambda < 1$ and $\Gtt$ for $\lambda \ne 0$.
Also indicated are the affine shear-elasticity $\muA(\lambda) \approx 34.2$ (dashed line) 
and the stress-stress fluctuation $\muTT$ which is seen to decay linearly (bold dash-dotted line)
according to eq.~(\ref{eq_keystat}) down to $\muTT|_{\lambda=1} \approx 18$. 
\label{fig_muF_L}
}
\end{figure}

\paragraph*{Strain-stress correlations.}
Since the shear modulus $\Geq$ is finite, this implies that shear strain and shear stress 
fluctuations must be correlated \cite{WXP13}. These correlations are addressed in fig.~\ref{fig_tau_gam}.
The main panel presents the pre-averaged stress $\tau(x)$ for an instantaneous strain 
$x = \gamhat - \gamext$ for several $\lambda$ \cite{foot_preaverage}. 
Irrespective of $x$ or $\lambda$
all data collapse on the linear slope indicated by the bold line with $\Geq=16.3$. 
Naturally, it follows from fig.~\ref{fig_histoL} that the statistics must strongly
decrease for $x^2 \gg \muGT \equiv \beta V \langle \delta \gamhat \delta \tauhat \rangle$, 
i.e. a larger linear-slope window is visible in fig.~\ref{fig_tau_gam} for smaller $\lambda$.
Panel (b) of fig.~\ref{fig_tau_gam} presents two scatter plots of $(\gamhat,\tauhat)$
for $\lambda=-1.23$ ($\Gext=-9$). The small circles correspond to a time series of 
100 subsequent and, hence, strongly correlated configurations with $\delta t = \dtMD$.
On these time scales the configuration has no time to relax the affine displacements
imposed by the shear-barostat. As emphasized by the dashed slope this leads
to a linear stress-strain relation with a coefficient $\muA = 34.2$.
Other short-time sequences yield similar, but horizontally shifted linear slopes (not shown).
A $\delta t$-independent representation of the static strain-stress correlations for large times 
is obtained if $(\gamhat,\tauhat)$ is indicated for 100 configurations with a time interval 
$\delta t = 10^2$ (squares). 
The scatter plot of the latter data is already in nice agreement 
with a coefficient $\Geq=16.3$ (bold line).
Sampling over {\em all} data tuples $(\gamhat,\tauhat)$ of a given $\lambda$-ensemble
one verifies that the shear modulus $\Geq$ is accurately determined using
the linear regression coefficient 
\begin{equation}
\Ggt \equiv \frac{\la \delta \gamhat \delta \tauhat \ra}{\la  \delta \gamhat^2 \ra} = \frac{\muGT}{\muGG}
\mbox{ for } \lambda < 1.
\label{eq_Ggtgen}
\end{equation}
Since, as shown in sect.~\ref{theo_stat}, $\muGG = (1-\lambda)/\Geq$ and $\muGT = 1-\lambda$, 
this ratio must yield $\Geq$ for all $\lambda < 1$.
Confirming $\Geq=16.3$ the values of $\Ggt$ for different $\lambda$ are 
identical to $\Ggg$ as shown in fig.~\ref{fig_muF_L}.

\paragraph*{Stress-stress fluctuations.}
The stress-stress fluctuations $\muTT \equiv \beta V \langle \delta \tauhat^2 \rangle$ 
presented in fig.~\ref{fig_muF_L} (circles) decrease linearly 
with $\lambda$ as stated in the Introduction, eq.~(\ref{eq_keystat}). Being a simple mean  
$\muA$ (small squares) is found to be strictly $\lambda$-independent. 
As expected, $\muTT \to \muA$ for $\lambda \to 0$.
Using eq.~(\ref{eq_keystat}) the generalized stress-fluctuation formula for $\Geq$ reads 
\begin{equation}
\Gtt \equiv \frac{\muA-\muTT}{1-(\muA-\muTT)/\Gext} \mbox{ for } \lambda \ne 0.
\label{eq_Gttgen}
\end{equation}
For $\lambda \to 1$, i.e. $\Gext \to \infty$, this reduces to eq.~(\ref{eq_Gtt}). 
Equation~(\ref{eq_Gttgen}) thus generalizes the stress-fluctuation formula 
for the $\NVgT$-ensemble 
\cite{FrenkelSmitBook,Hoover69,Barrat88,Lutsko89,Barrat13,WXP13,WXB15,WXBB15}
to general $\lambda$.
As seen from fig.~\ref{fig_muF_L} (diamonds), it is thus possible to determine $\Geq$ 
from $\muA$ and $\muTT$ for all $\lambda \ne 0$.

\subsection{Dynamics: Strain-strain correlations}
\label{simu_dyna_GG}

\begin{figure}[t]
\centerline{\resizebox{1.0\columnwidth}{!}{\includegraphics*{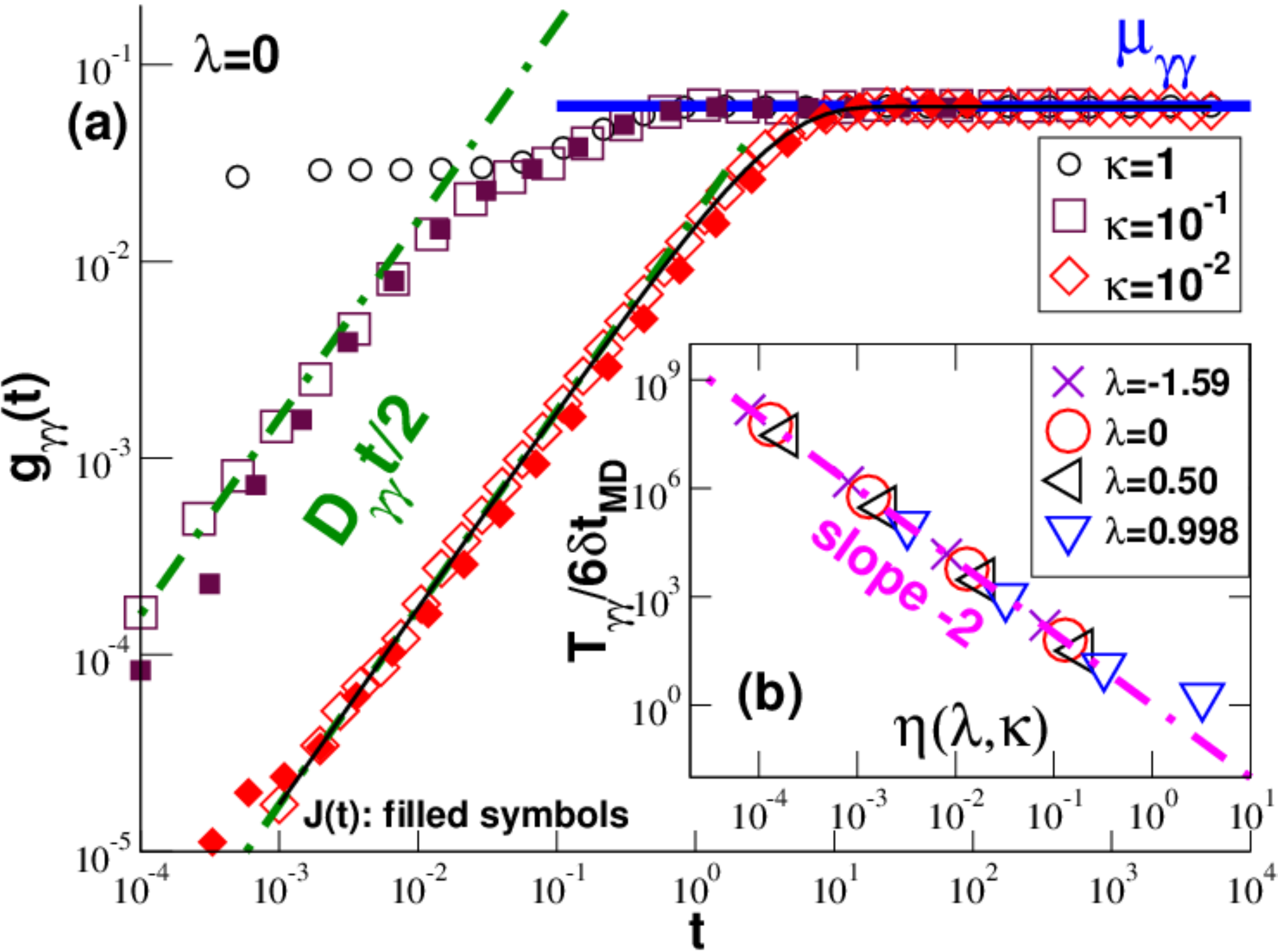}}}
\caption{Diffusion of instantaneous strain $\gamhat(t)$: 
{\bf (a)} 
MSD $\gGGt$ for $\lambda=0$ (open symbols) compared to the explicitly 
computed creep compliance $J(t) = \langle \delta \gamhat(t) \rangle/\delta \tauext$ 
for a step stress $\delta \tauext = 0.01$ (filled symbols).
The dash-dotted lines indicate the regime used 
to determine the diffusion coefficient $\Dgg$,
the bold horizontal line the expected asymptotic limit $\muGG$
and the thin solid line the Kelvin-Voigt model, eq.~(\ref{eq_KelvinVoigt}), for $\kappa=10^{-2}$.
{\bf (b)}
Double-logarithmic representation of the rescaled crossover time $\Tgg/6\dtMD$ {\em vs.} $\eta$. 
The dash-dotted line indicates the expected power law, eq.~(\ref{eq_Tgg_FDL}).
\label{fig_gGGt}
}
\end{figure}

\paragraph*{Strain-strain MSD.}
The strain histograms $p(x)$ for different $\lambda$ and their second moments
shown, respectively, in fig.~\ref{fig_histoL} and fig.~\ref{fig_dgam} have been 
rapidly sampled by averaging over short time series from trajectories of length 
$\ttraj=10^4$.  This is demonstrated in panel (c) of fig.~\ref{fig_dgam} 
presenting the MSD $\gGGt$ of the instantaneous strain $\gamhat(t)$ for $\kappa=10^{-1}$. 
The statistics is improved by performing a gliding average
over the $10^8$ data tuples sampled \cite{AllenTildesleyBook}.
It is seen that $\gGGt$ converges rapidly after a time $\Tgg$ of order unity to its asymptotic limit
$\muGG(\lambda)$ indicated by horizontal lines. 
Free diffusion is observed for short times, i.e. $\gGGt \approx \Dgg t /2$ for $t \ll \Tgg$ 
(dash-dotted line), and this essentially irrespective of the ensemble in agreement with 
eq.~(\ref{eq_Dgg_FDL}). 
Note that free strain diffusion is observed for all $\lambda$ and sufficiently 
small $\kappa$ if the parameter $\eta(\lambda,\kappa) \ll 1$.
As shown in the main panel of fig.~\ref{fig_gGGt} for different $\kappa$ and $\lambda=0$
(open symbols), this behavior may be used to determine first the short-time diffusion coefficient 
$\Dgg$ for systems with sufficiently small $\eta$ and then using eq.~(\ref{eq_TabDab}) 
the crossover time $\Tgg$ of the strain fluctuations. 
The rescaled times $\Tgg/6 \dtMD$ are represented in panel (b) of fig.~\ref{fig_gGGt} 
as a function of the scaling variable $\eta$ for several $\lambda$. The dash-dotted line 
indicates the power-law slope, eq.~(\ref{eq_Tgg_FDL}).
A perfect data collapse on this line is observed for all $\eta \ll 1$.
A successful data collapse can be also obtained for $\gGGt$ over a broad range of $\lambda$ and 
$\kappa$ by plotting $\gGGt/\muGG$ as a function of the reduced time $x= t/\Tgg$ (not shown).
As seen by the thin solid line in the main panel of fig.~\ref{fig_gGGt}, 
one verifies that the scaling function $f(x) = \gGGt/\muGG$ is given by $f(x)=1-\exp(-x)$.
This is, of course, consistent with an exponentially decaying strain-strain
relaxation function $\cGGt = \muGG \exp(-x)$ as we have also checked directly.

\paragraph*{Creep compliance $J(t)$.}
As discussed in sect.~\ref{theo_dyna_GtJt}, a MSD $\gGGt$ computed in the $\NVtT$-ensemble
corresponds to a creep compliance $J(t)= \langle \delta \gamhat(t)\rangle/\delta \tauext$ 
measured by imposing at $t=0$ a step stress increment $|\delta \tauext| \ll 1$ to the external
shear stress $\tauext$ applied to the system.
Using $\delta \tauext =0.01$ and averaging over 100 configurations this yields the
data shown by the filled symbols in panel (a) of fig.~\ref{fig_gGGt} for two values of $\kappa$.
An excellent data collapse $\gGGt \approx J(t)$ is observed.
While it is natural to use the \NVtT-ensemble for obtaining the creep compliance from the equilibrium 
fluctuations, it is worthwhile to note that due to the scaling $\gGGt / \muGG =  f(x)$, 
$J(t)$ may be also determined from other $\lambda$-values.
\begin{figure}[t]
\centerline{\resizebox{1.0\columnwidth}{!}{\includegraphics*{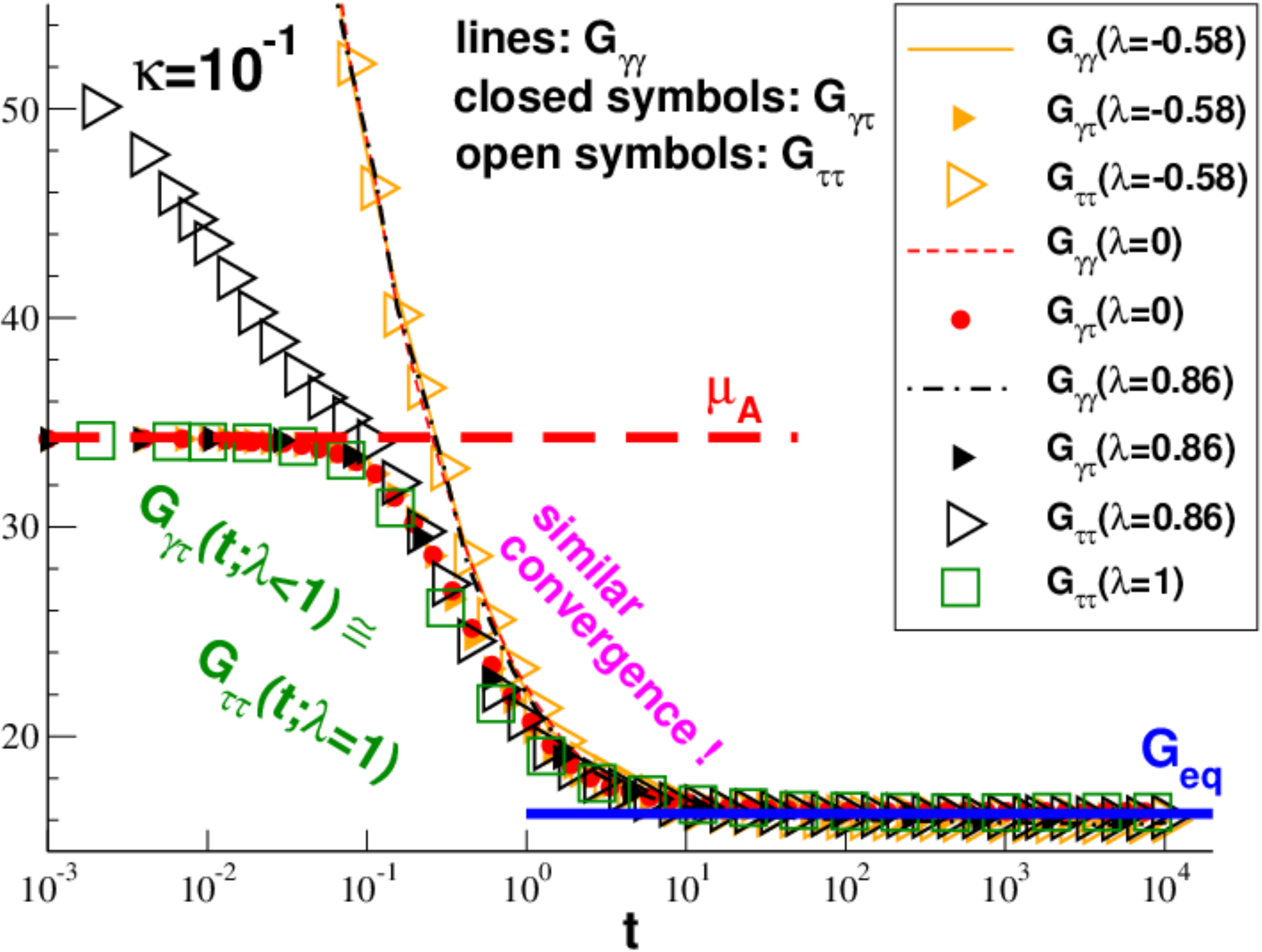}}}
\caption{Determination of equilibrium shear modulus $\Geq$ as a function of
the sampling time $t$ for $\Gggt$ (thin lines), $\Ggtt$ (filled symbols) and $\Gttt$ (large open symbols)
for $\kappa=10^{-1}$.
All fluctuation formulae are equivalent for $t > 1$ and converge similarly to $\Geq$ for large times.
Note that $\Ggtt$ for all $\lambda < 1$ is everywhere identical to $\Gttt$ for $\lambda=1$.
\label{fig_Geq_t}
}
\end{figure}

\paragraph*{Time-dependent strain-strain fluctuations.}
As expected from sect.~\ref{theo_dyna_genrel}, the strain-strain fluctuations $\muGGt$ computed 
according to eq.~(\ref{eq_muABn_two}) by gliding average from a finite time series 
increase monotonously with time $t$ from zero (if only one configuration is sampled) 
to the asymptotic thermodynamic limit $\muGG$ (not shown).
Note that one determines first the strain fluctuations in an interval $[t_0,t_1=t_0+t]$ and 
averages then over all possible $t_0$.
If the estimate $\Ggg$, eq.~(\ref{eq_Ggggen}), of the modulus $\Geq$ is 
determined using $\muGGt$ instead of $\muGG$, it becomes a time dependent quantity
called $\Gggt$. As can be seen from fig.~\ref{fig_Geq_t} for three different $\lambda$
values, $\Gggt$ thus becomes a monotonously decreasing function of time (thin lines).
Interestingly, $\Gggt$ appears not to depend on $\lambda$.
We note finally that using either the large-time asymptotics eq.~(\ref{eq_muABt_asymp})
or, more conveniently, the Debye relation eq.~(\ref{eq_Debye}), one may determine
the relaxation time $\thetaGG$ of the strain-strain fluctuations $\muGGt$. 
One verifies that $\thetaGG \approx \Tgg$ 
holds as expected for an exponentially decaying correlation function (sect.~\ref{theo_dyna_model}).

\subsection{Dynamics: Strain-stress correlations}
\label{simu_dyna_GT}

\begin{figure}[t]
\centerline{\resizebox{1.0\columnwidth}{!}{\includegraphics*{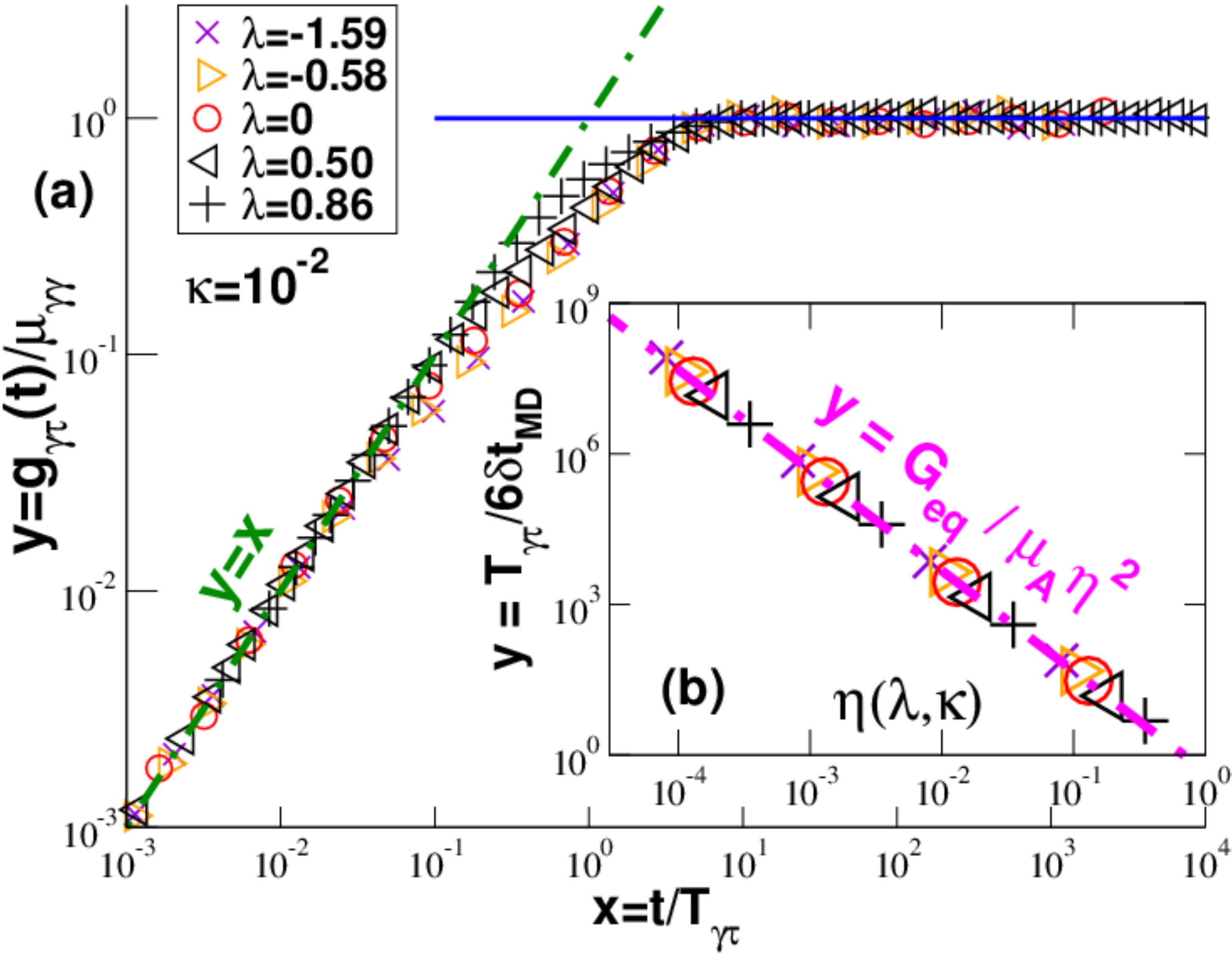}}}
\caption{Scaling of strain-stress MSD $\gGTt$ for different $\lambda$:
{\bf (a)} 
scaling collapse of $y=\gGTt/\muGT$ {\em vs.} $x=t/\Tgt$ for $\kappa=10^{-2}$
with the dash-dotted line indicating the diffusive regime and the bold solid line
the long-time limit,
{\bf (b)}
crossover time $\Tgt$, eq.~(\ref{eq_TabDab}), rescaled as $\Tgt/6\dtMD$ {\em vs.} 
the scaling variable $\eta$ using the same symbols as in the main panel.
The dash-dotted line indicates the prediction eq.~(\ref{eq_TgtTgg}).
\label{fig_gGTt}
}
\end{figure}

\paragraph*{Strain-stress MSD.}
The scaling of the strain-stress MSD $\gGTt$ is investigated in fig.~\ref{fig_gGTt}. 
The relaxation time $\Tgt$ presented in panel (b) has been determined, as before for $\Tgg$,
by matching the short-time diffusive regime with the long-time limit $\gGTt \to \muGT$. 
As indicated by the dash-dotted line, the data are well described by
\begin{equation}
\Tgt = \frac{\Geq}{\muA} \Tgg \sim 1/\kappa^2
\mbox{ for } \eta \ll 1.
\label{eq_TgtTgg}
\end{equation}
That this scaling must hold can be seen by replacing in the definition of $\gGTt$
the stress fluctuation $\delta \tauhat(t)$ by $\muA \delta \gamhat(t)$ which
implies $\Dgt = \muA \Dgg$ for the diffusion coefficients and in turn eq.~(\ref{eq_TgtTgg})
using again eq.~(\ref{eq_TabDab}).
We stress that in agreement with the inset of fig.~\ref{fig_tau_gam} the system has not enough
time to relax the affine strain imposed by the shear-barostat for the short times
$t \ll \tauA \approx 0.1$ considered here. It is thus $\muA$ and not $\Geq$ 
which has to be used as the linear slope coefficient.
The main panel (a) presents the collapse of $y = \gGTt/\muGT$ as a function of $x = t/\Tgt$
for $\kappa=10^{-2}$. Minor deviations are visible for $x \approx 1$.
The quality of the collapse improves by decreasing $\eta(\lambda,\kappa)$,
whereas the crossover becomes more sudden (even a hump may occur) if $\eta$ is too large (not shown).

\paragraph*{Time-dependent correlation coefficient $\Ggtt$.}
Since the strain-shear correlations are dominated for $t \ll \tauA$ by the {\em affine} shearing,
one expects $\Ggtt$ computed using the ratio of $\muGTt$ and $\muGGt$ to be similar to $\muA$ for $t \to 0$.
As one sees from $\Ggtt$ presented in fig.~\ref{fig_Geq_t} (filled symbols),
this is indeed the case. Interestingly, 
$\Ggtt$ is seen to be independent of $\lambda$ for {\em all} times $t$,
i.e. the $\lambda$-dependences of $\muGTt$ and $\muGGt$ do cancel for all times
just as they cancel for the ratio of the equilibrium moments $\muGT/\muGG$.

\subsection{Dynamics: Stress-stress correlations}
\label{simu_dyna_TT}

\begin{figure}[t]
\centerline{\resizebox{1.0\columnwidth}{!}{\includegraphics*{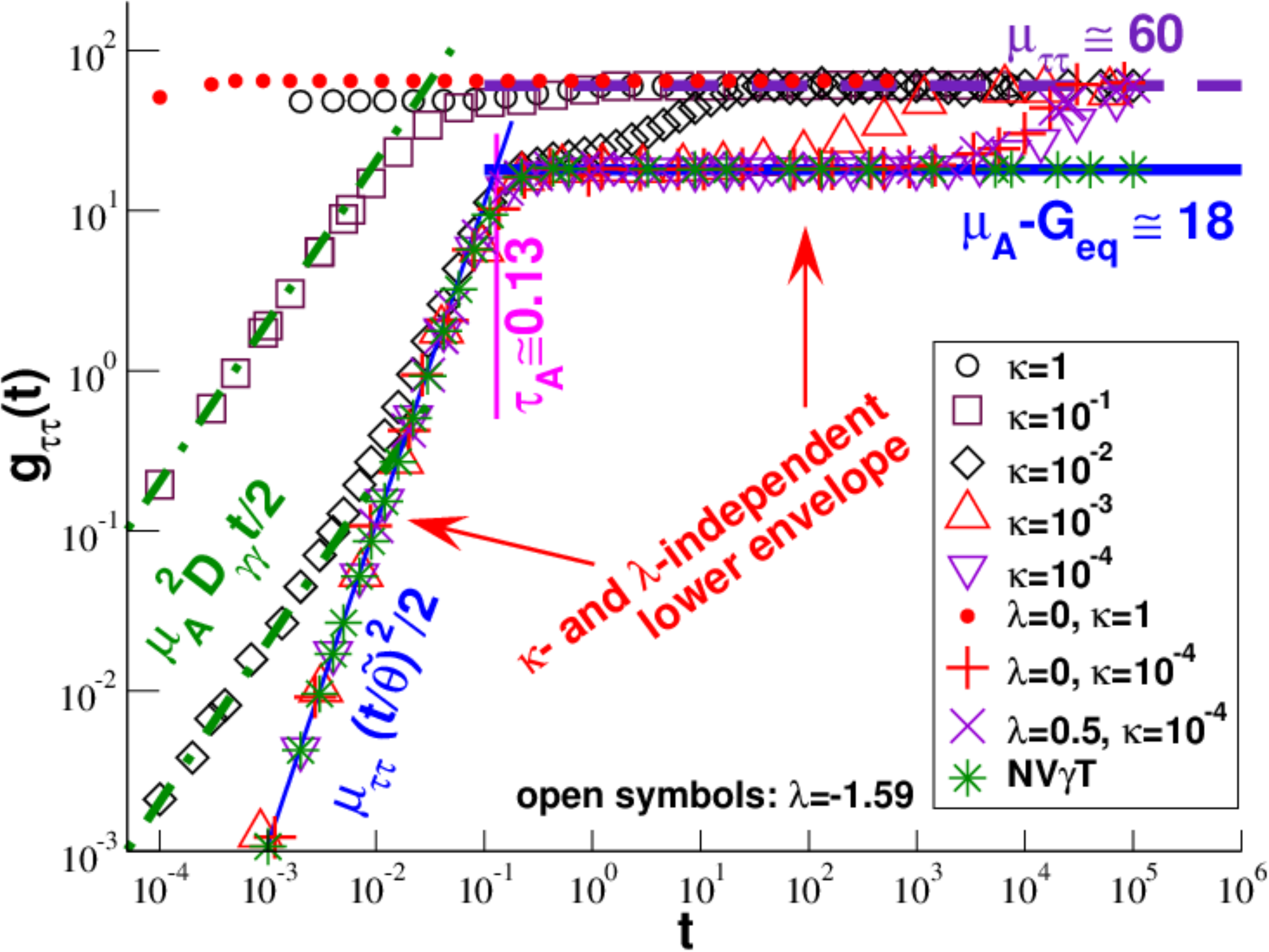}}}
\caption{Stress-stress MSD $\gTTt$ focusing on $\lambda=-1.59$ comparing different $\kappa$ (open symbols).
Additionally, we present $\lambda=0$ for $\kappa=1$ (small filled circles) and $\kappa=10^{-4}$,
$\lambda=0.5$ for $\kappa=10^{-4}$ and the $\NVgT$-ensemble (stars).
The dot-dashed lines indicate the free diffusive behavior for intermediate $\eta$,
the thin solid line the analytic short-time limit for small $\eta$,
the horizontal dashed line the thermodynamic limit $\muTT$ for $\lambda=-1.59$
and the bold solid line the expected intermediate plateau, 
eq.~(\ref{eq_gTTt_plateau}). 
Note that for sufficiently small $\kappa$ all data approach the $\kappa$- and $\lambda$-independent
lower envelope indicated by the solid lines in agreement with the fundamental scaling postulate 
eq.~(\ref{eq_MSDscaling}).
\label{fig_gTTt}
}
\end{figure}

\paragraph*{Introduction.}
We turn finally to the more intricate characterization of the stress-stress correlations 
as a function of our two operational parameters $\lambda$ and $\kappa$. 
We focus again on the regime with $\eta(\lambda,\kappa) \ll 1$. As above we begin by 
discussing the MSD $\gTTt$. 
Since according to eq.~(\ref{eq_gABt_cABt}) 
\begin{equation}
\gTTt = C_\mathrm{\tau\tau}(0) - \cTTt \mbox{ with } C_\mathrm{\tau\tau}(0) = \muTT,
\label{eq_gTTt_cTTt}
\end{equation}
the MSD $\gTTt$ and the correlation function $\cTTt$
contain in principal the same information. From the presentational point of view $\gTTt$ 
has the advantage that the short-time power-law behavior of the correlations can be made manifest
using a double-logarithmic plot as shown in fig.~\ref{fig_gTTt}. 
We describe then the large time behavior of the correlation function $\cTTt$
and demonstrate that $\tstar$ is given by $\Tgg$. Finally, we turn to the time-dependent 
stress fluctuation $\muTTt$ which is used to determine the stress-stress relaxation time $\thetaTT$.

\paragraph*{Stress-stress MSD.}
Let us concentrate first on the data for $\lambda=-1.59$ presented by open symbols in fig.~\ref{fig_gTTt}.
If $\kappa$ is large, all internal dynamics is destroyed and 
we find immediately $\gTTt \approx \muTT$ as seen for $\kappa=1$ (small circles).
For smaller, but not too small $\kappa$ one observes a free-diffusion regime with 
$\gTTt = \Dtt t/2$ as shown by the bold dash-dotted lines on the left.
As above for $\gGTt$ in sect.~\ref{simu_dyna_GT}, $\delta \tauhat(t)$ in the definition
of $\gTTt$ can be replaced by $\muA \delta \gamhat(t)$.
This argument yields the diffusion coefficient $\Dtt = \muA^2 \Dgg$ successfully used in the plot
for $\kappa=10^{-1}$ and $\kappa=10^{-2}$.
If $\kappa$ is further reduced, $\gTTt$ becomes $\kappa$-independent for short times.
As seen for $\kappa=10^{-3}$ and $10^{-4}$ the dynamics becomes ``deterministic"
in the sense that 
\begin{equation}
\gTTt = \frac{\muTT}{2} \ (t/\thetatilde)^2 \ \sim \kappa^0 \lambda^0 \mbox{ for } t \ll \tauA  
\label{eq_gTTt_determinist}
\end{equation}
with $\thetatilde \approx 0.16$ as shown by the thin solid line. 
Albeit $\muTT$ depends on $\lambda$, the ratio $\muTT/\thetatilde^2$ does not, in agreement
with the fundamental postulate eq.~(\ref{eq_MSDscaling}).
This implies that $\thetatilde$ depends somewhat on $\lambda$.
The $t^2$-scaling in eq.~(\ref{eq_gTTt_determinist}) is expected \cite{WXBB15} since the associated correlation 
function $\cTTt$ must be a continuous and symmetric function at $t=0$ which, moreover, should be an 
{\em analytic} function if barostat and thermostat effects can be ignored \cite{foot_CtGauss}.
This implies that $\gTTt$ must be an even expansion in terms of $t^2$
which to leading order leads to eq.~(\ref{eq_gTTt_determinist}).
Since there are several dynamical regimes it is not meaningful to determine a crossover time $\Ttt$
using eq.~(\ref{eq_TabDab}) as before for $\Tgg$ and $\Tgt$ over the full range of $\lambda$ and $\kappa$.
We shall see at the end of this section how a longest relaxation time $\thetaTT$ 
for the stress-stress correlations might be obtained.
For times $t \gg \tauA$ an intermediate plateau appears (bold solid line)
as predicted by eq.~(\ref{eq_cTTt_plateau}).
As can be seen for four different $\lambda$ values this intermediate 
plateau does {\em not} depend on the ensemble provided that $\kappa$ is sufficiently weak. 
(As one expects from fig.~\ref{fig_Aeta} deviations from this asymptotic limit arise again
for large $\kappa$ if $\eta(\lambda,\kappa) \gg 0.1$.)
We emphasize that the two solid lines indicated in fig.~\ref{fig_gTTt} give the lower envelope for 
all parameters $\lambda$ and $\kappa$ we have simulated. 
The MSD $\gTTt$ behaves thus as an ensemble-independent simple average
as we have argued in sect.~\ref{theo_dyna_lambda}.
One may operationally define the crossover time $\tauA$ by matching eq.~(\ref{eq_gTTt_determinist}) 
and eq.~(\ref{eq_gTTt_plateau}). This yields
\begin{equation}
\tauA = \thetatilde \ \sqrt{2 \ \frac{1-\Geq/\muA}{1-\lambda \Geq/\muA}} \approx 0.13
\label{eq_tauA}
\end{equation}
as indicated by the vertical line. Note that $\tauA = \thetatilde \sqrt{2}$ for $\lambda=1$.
We stress that while $\thetatilde$ is a function of $\lambda$, 
$\tauA$ is an ensemble-independent constant for the given elastic network.
(Moreover, it can be shown that it barely depends on the friction constant $\zeta$ of the Langevin 
thermostat \cite{WXBB15}.) Using eq.~(\ref{eq_tauA}) one may reformulate 
eq.~(\ref{eq_gTTt_determinist}) in a manifest $\lambda$-independent form:
$\gTTt = (\muA-\Geq) \ (t/\tauA)^2$.
The MSD ultimately approaches $\muTT(\lambda)$ for even larger times 
$t \gg \tstar$ as indicated by the dashed horizontal line for $\lambda=-1.59$. 
Remembering that $\muTT = \muA-\Geq$ for $\lambda=1$ it is clear 
that the \NVgT-data stays constant for all times.

\begin{figure}[t]
\centerline{\resizebox{1.0\columnwidth}{!}{\includegraphics*{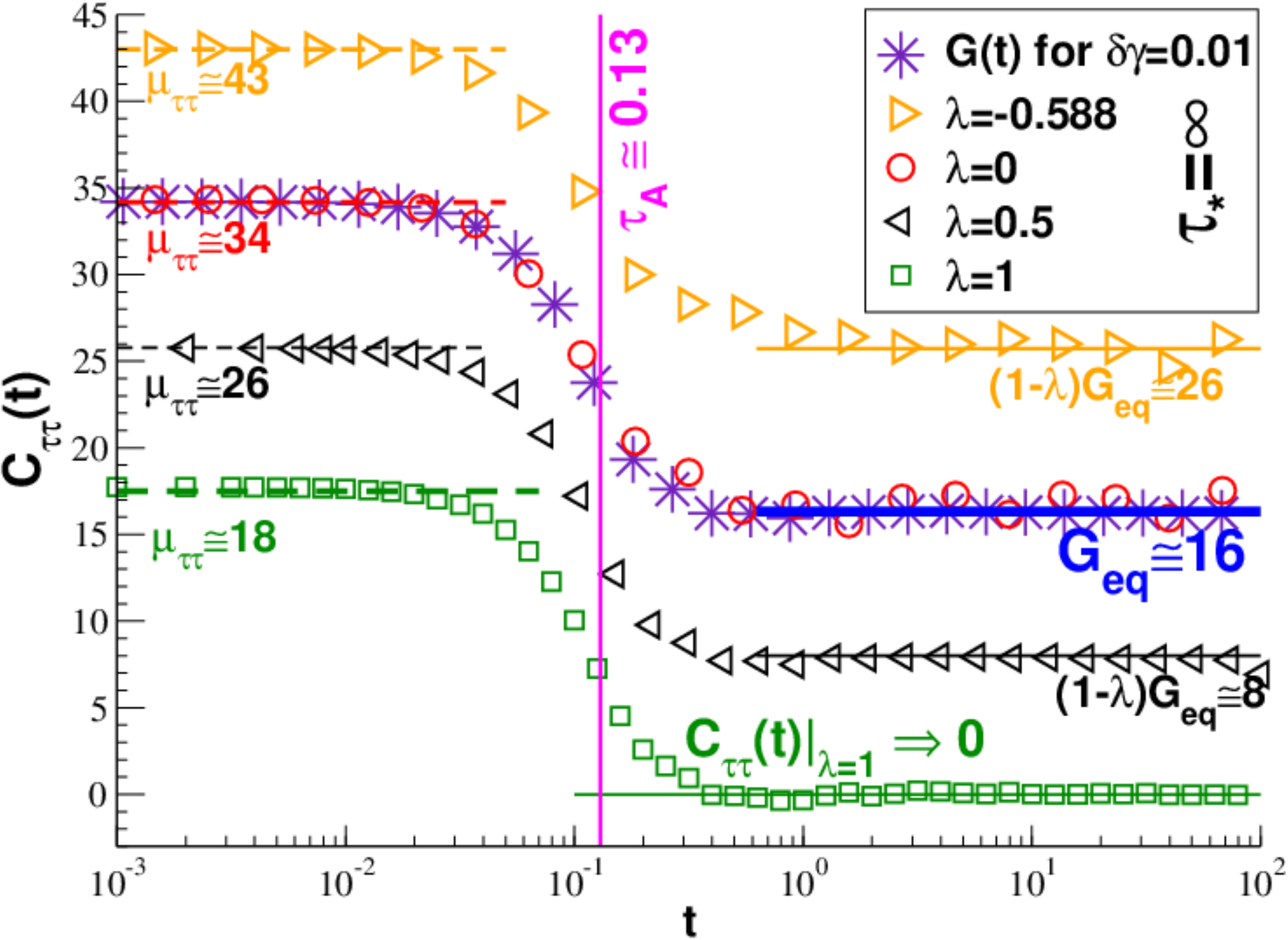}}}
\caption{Relaxation modulus $G(t)$ obtained from the stress response to an applied
strain $\delta \gamma =0.01$ (stars) compared to the equilibrium correlation function 
$\cTTtlam$ for several $\lambda$ obtained by averaging over 1000 quenched-strain configurations.
\label{fig_cTTt}
}
\end{figure}

\begin{figure}[t]
\centerline{\resizebox{1.0\columnwidth}{!}{\includegraphics*{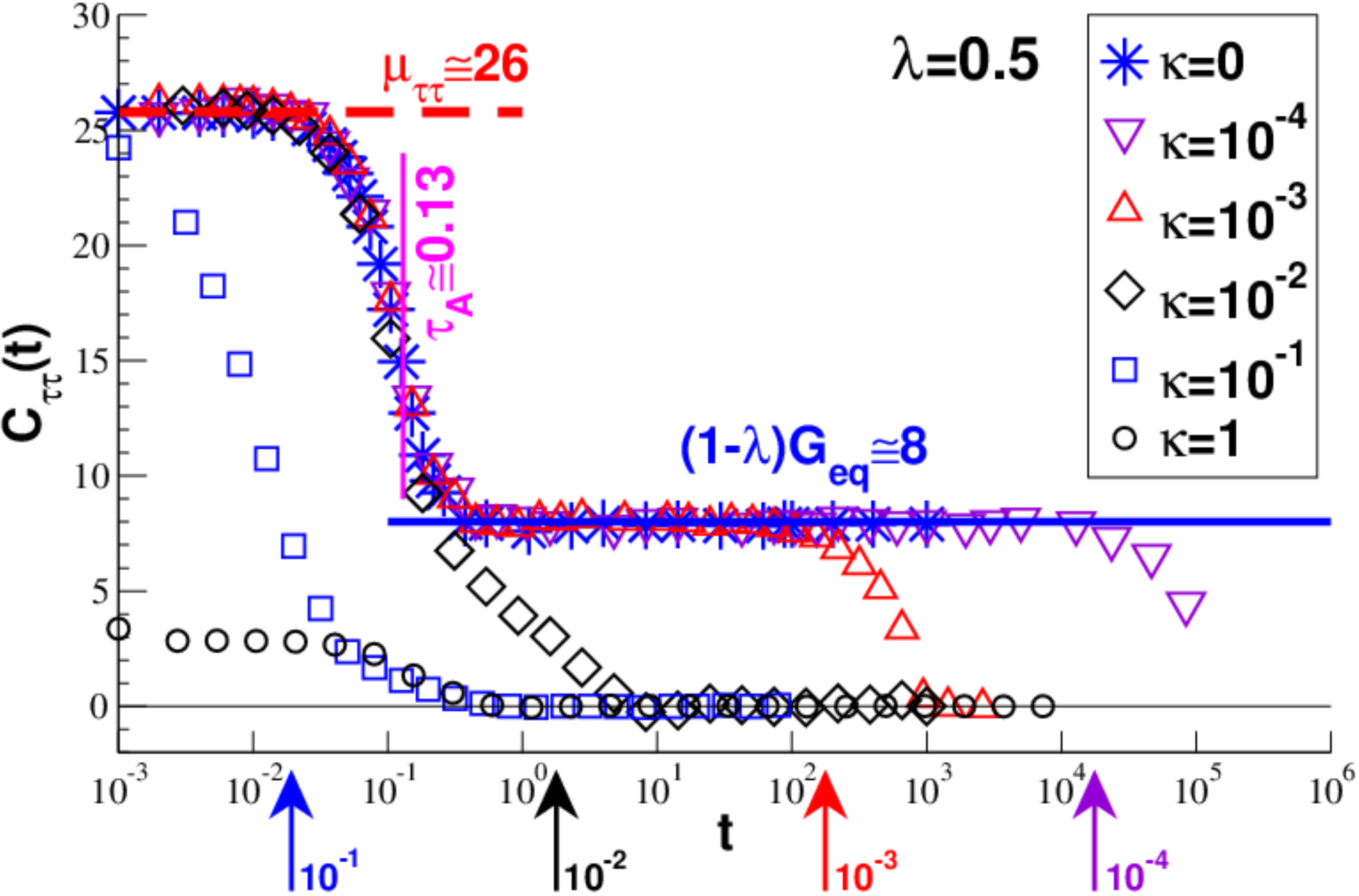}}}
\caption{Effect of the barostat parameter $\kappa$ for $\cTTt$ with $\lambda=0.5$.
The large stars correspond to the quenched $\gamma$-ensemble ($\tstar = \infty$)
and all other symbols to a switched-on barostat of finite $\kappa$.
The vertical arrows indicate $\Tgg$ for $\kappa= 10^{-1}, 10^{-2}, 10^{-3}$ and $10^{-4}$
confirming that roughly $\Tgg \approx \tstar$. A much longer run is warranted for $\kappa=10^{-4}$.
\label{fig_cTTt_Gxymax}
}
\end{figure}

\paragraph*{Relaxation modulus.}
The large-$t$ scaling is further addressed in fig.~\ref{fig_cTTt} and fig.~\ref{fig_cTTt_Gxymax} 
presenting the stress-stress correlation function $\cTTt$.
The most central result of this work stated by eq.~(\ref{eq_keydyna}) is demonstrated in fig.~\ref{fig_cTTt}. 
We compare the directly measured relaxation modulus $G(t)$ 
with the equilibrium correlation function $\cTTt$ assuming an asymptotically slow barostat $(\tstar=\infty$). 
Averaging over 1000 independent configurations from an $\NVgT$-ensemble with $\gamma=0$ for $t < 0$,
the relaxation modulus $G(t) = \langle \tauhat(t) \rangle/\delta \gamma$ measures 
the stress response after a (canonical and affine) shear strain $\delta \gamma = 0.01$ 
was been applied at $t=0$ (stars).
The correlation functions $\cTTt$ for $\lambda < 1$ have been obtained by averaging
over 1000 equilibrated configurations from a $\lambda$-ensemble as indicated.
Switching off the shear-barostat the strain $\gamhat$ of each configuration
is quenched. Note that $C_{\tau\tau}(0)=\muTT$ holds due to the ensemble averaging. 
As expected from eq.~(\ref{eq_cTTt_plateau}), $\cTTt$ decreases monotonously from $\muTT$ down to 
$(1-\lambda) \Geq$ for $t \gg \tauA$. 
This corresponds to the $\lambda$-independent intermediate plateau $\gTTt=\muA-\Geq$
in fig.~\ref{fig_gTTt}.
It is seen that $G(t) = \cTTt$ for $\lambda=0$ (large circles) and that for all $\lambda$ 
one may obtain $G(t)$ by vertically shifting the correlation functions
$\cTTtlam  \rightarrow \cTTtlam + \lambda \Geq = G(t)$
using the already determined shear modulus $\Geq$.
Note also that $\cTTt|_{\lambda=1} \to 0$ for large times while all other
$\cTTt$ remain finite. As already emphasized at the end of sect.~\ref{theo_dyna_GtJt},
it is thus impossible to obtain the modulus $\Geq$ {\em solely} from  $\cTTt|_{\lambda=1}$. 
Please note that the observed short-time value $\muTT \approx 18$ (bold dashed line)
for $\lambda=1$ is quite different from the equilibrium modulus $\Geq \approx 16$
(bold solid line).

\paragraph*{Finite-$\kappa$ effects.}
Focusing on one ensemble with $\lambda=0.5$ 
we investigate in fig.~\ref{fig_cTTt_Gxymax} the effect of a switched-on shear barostat.
The stars refer to data already seen in fig.~\ref{fig_cTTt}
using the quenched-strain ensemble ($\kappa=0$, $\tstar=\infty$).
As one expects, the decay of $\cTTt$ becomes systematically slower with decreasing $\kappa$.
For large $\kappa > 10^{-2}$ we have computed $\cTTt$ by gliding averaging
over a single trajectory of $\ttraj=10^4$. For smaller $\kappa$ it becomes increasingly
harder to have a sufficiently long trajectory of $\{\tauhat_k\}$ probing the phase space, 
i.e. for which $C_{\tau\tau}(0)=\muTT$, 
and to store at the same time the shear stresses with a sufficiently fine time resolution.
The data for $\kappa \le 10^{-2}$ have thus been obtained by using an ensemble
of $100$ configurations of different $\gamhat$ and averaging over the trajectories
obtained for each configuration using a finite $\kappa$.
As expected from eq.~(\ref{eq_cTTt_plateau}), for sufficiently small $\kappa$ all correlation functions
have an intermediate plateau for $\tauA \ll t \ll \tstar$ (bold solid line).
That the barostat is not completely switched off can be seen from the final decay.
The vertical arrows indicate the crossover times $\Tgg$ for the four smallest $\kappa$.
As one expects, this time scale appears to coincide roughly (up to a constant prefactor) 
with the upper limit $\tstar$ of the respective plateau. In other words, the plateau
disappears when the ergodicity breaking associated with the averaging over a quenched-strain 
ensemble is lifted by the shear-barostat.

\paragraph*{Time-dependent stress-fluctuation formula.}
Let us return to fig.~\ref{fig_Geq_t}.
According to the generalized stress-fluctuation formula $\Gtt$, eq.~(\ref{eq_Gttgen}),
the shear modulus $\Geq$ may be obtained from the affine shear-elasticity $\muA$
and the equilibrium stress-stress fluctuation $\muTT$ for $\lambda \le 1$ and $\lambda \ne 0$
(fig.~\ref{fig_muF_L}). 
If instead of $\muTT$ a time-averaged fluctuation $\muTTt$ is used,
this leads again to a time-dependent monotonously decreasing estimate $\Gttt$
as seen from the open symbols in fig.~\ref{fig_Geq_t}. 
No shear-barostat is of course applied for the $\NVgT$-ensemble indicated by the large squares.
We emphasize the remarkable finding that within numerical accuracy
\begin{equation}
\Gttt|_{\lambda=1} \approx \Ggtt|_{\lambda<1}
\label{eq_striking}
\end{equation}
for all times and a broad range of $\kappa$ for $\Ggtt$.
As seen for $\lambda= - 0.58$, for small $\lambda$ and not too short times 
$\Gttt$ and $\Gggt$ are generally similar.
Since $\muTTt$ must vanish for small times, this implies 
that $\Gttt$ must become finite
\begin{equation}
\Gttt \to \frac{\muA}{1-\muA/\Gext} \mbox{ for } t \to 0
\label{eq_Gtt_smallt}
\end{equation}
and, hence, $\Gttt \to \muA$ for $\Gext \gg \muA$.
This limit is indicated by the bold dashed line for $\lambda=1$. 
\begin{figure}[t]
\centerline{\resizebox{1.0\columnwidth}{!}{\includegraphics*{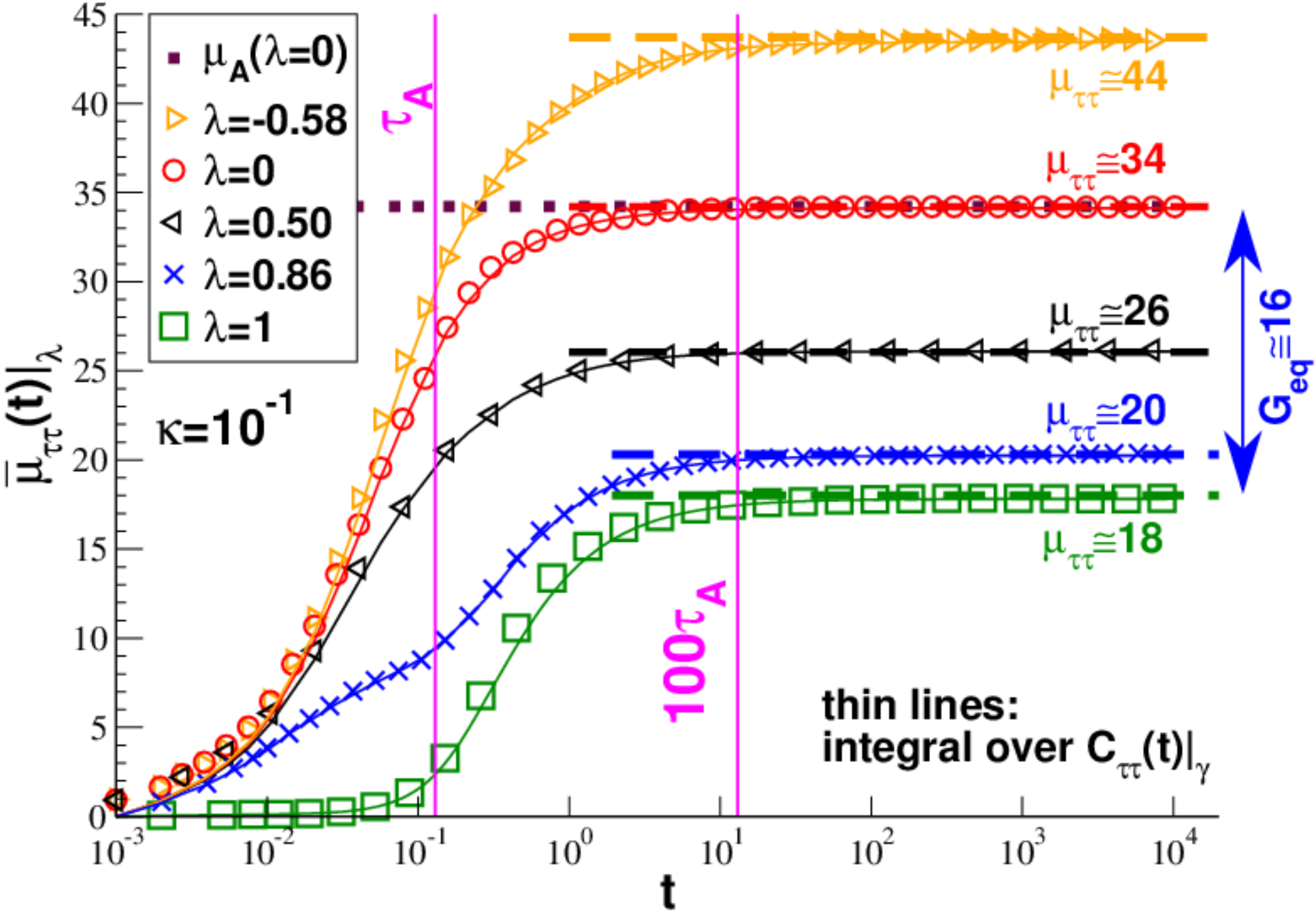}}}
\caption{Shear-stress fluctuation $\muTTtlam$ as a function of the sampling time.
Also indicated is the affine shear-elasticity $\muA$ obtained for the $\NVtT$-ensemble 
(small filled squares) which is seen to immediately converge to the long-time limit. 
The thin lines correspond to the integrated shear-stress auto-correlation function 
$\cTTtlam$ according to eq.~(\ref{eq_muABt_cABt}).
\label{fig_muF_t}
}
\end{figure}

\begin{figure}[t]
\centerline{\resizebox{1.0\columnwidth}{!}{\includegraphics*{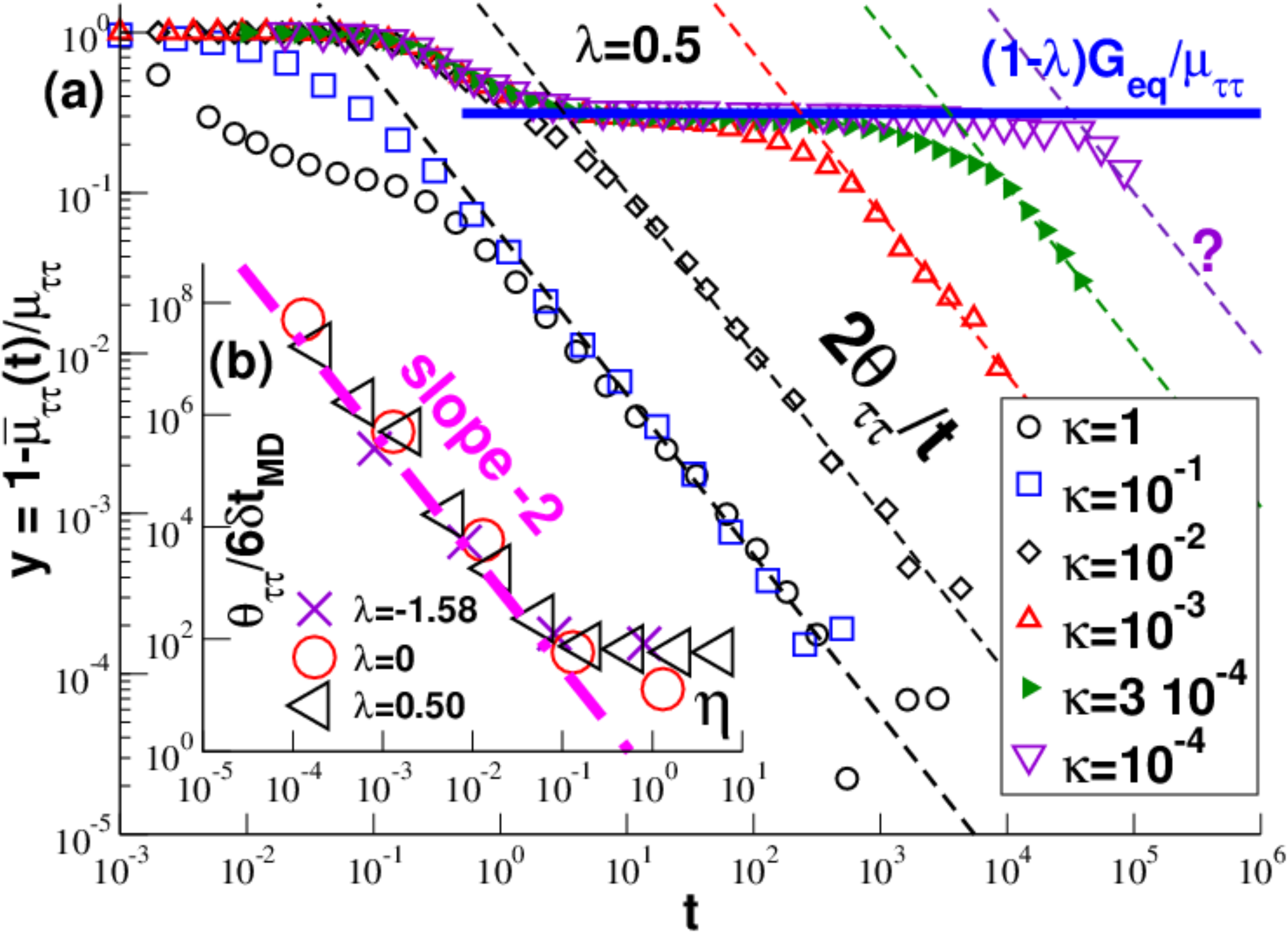}}}
\caption{Stress-stress relaxation time $\thetaTT$:
{\bf (a)} Determination of $\thetaTT$ from the decay of $y=1-\muTTt/\muTT$ according to
eq.~(\ref{eq_muABt_asymp}) for $\lambda=0.5$ for different $\kappa$. The thin dashed lines
indicate eq.~(\ref{eq_muABt_asymp}), the bold horizontal line the expected intermediate plateau
$(1-\lambda)\Geq/\muTT$ for $10\tauA \ll t \ll \tstar$. 
{\bf (b)} Relaxation time $\thetaTT$ for three not too large $\lambda$ plotted
as $\thetaTT/6\dtMD$ {\em vs.} $\eta(\lambda,\kappa)$.
The dash-dotted line indicates the power law expected for $\eta \ll 1$
according to eq.~(\ref{eq_thetaTT}).
\label{fig_thetaTT}
}
\end{figure}

\paragraph*{Time-dependence of stress-stress fluctuations.}
The time-dependence of $\muTTt$ for several $\lambda$ is explicitly represented in fig.~\ref{fig_muF_t}
assuming $\kappa = 10^{-1}$ for $\lambda < 1$.
While the simple mean $\muA$ converges immediately with time, $\muTTt$ increases monotonously from zero 
to the large-time limit $\muTT$ (horizontal lines). 
As already pointed out in sect.~\ref{theo_dyna_genrel},
assuming time-translational invariance the time-dependence of $\muTTt$ 
is a consequence of the time-dependence of the $\cTTt$.
To emphasize this point we have integrated $\cTTt$ as
suggested by eq.~(\ref{eq_muABt_cABt}). As seen by the thin lines indicated
in fig.~\ref{fig_muF_t}, this leads to consistent results.
If one has thus characterized $\cTTt$, 
this implies $\muTTt$ and then in turn $\Gttt$.
Note that for all $\lambda$ the asymptotic limit is reached at about $100 \tauA$
which is of the same order as $\Tgg \approx 3.5$ for $\lambda=0$ and $\kappa=10^{-1}$.

\paragraph*{Relaxation time $\thetaTT$.}
The convergence of $\muTTt$ is systematically analyzed in fig.~\ref{fig_thetaTT}.
The main panel presents the dimensionless deviation $y = 1-\muTTt/\muTT$ for several
$\kappa$ and $\lambda=0.5$. 
As emphasized by the thin dashed lines, a correlation time $\thetaTT$ may be measured from 
the large-time $1/t$-decay following eq.~(\ref{eq_muABt_asymp}). 
(Unfortunately, the sampling time is yet insufficient for our smallest $\kappa$.)
With decreasing $\kappa$ the decay becomes systematically slower and a horizontal plateau 
with $y = (1- \lambda) \Geq/\muTT$ (bold solid line) appears in an intermediate time window. 
This plateau is expected from eq.~(\ref{eq_intplat}) and eq.~(\ref{eq_cTTt_plateau}).
Following the discussion in sect.~\ref{theo_dyna_genrel}, this implies 
that for sufficiently small $\kappa$ and not too large $\lambda$ the relaxation time $\thetaTT$ must scale 
(up to a prefactor of order unity) as
\begin{equation}
\thetaTT \approx 
\frac{(1- \lambda) \Geq}{\muTT} \ \tstar \sim (1-\lambda)^2/\kappa^2 \mbox{ for } \eta \ll 1
\label{eq_thetaTT}
\end{equation}
to leading order where we have used eq.~(\ref{eq_Tgg_FDL}) for $\tstar \approx \Tgg \sim 1/\kappa^2$.
That the $\kappa$-scaling holds can be seen in panel (b) of fig.~\ref{fig_thetaTT}.
(For $\lambda=0.5$ we have computed $\thetaTT$ also for the
additional values $\kappa = 3, 0.3, 0.03, 0.003$ and $0.0003$.)
The same representation as in the insets of figs.~\ref{fig_gGGt} and \ref{fig_gGTt}
has been chosen for comparison.
It should be stressed, however, that for $\thetaTT$
this is for two reasons not a rigorous scaling plot over the full 
range of operational parameters considered. Firstly, 
due to the $(1- \lambda) \Geq/\muTT$-factor in eq.~(\ref{eq_thetaTT}) there is an additional 
$\lambda$-dependence even for $\eta \ll 1$ and not too large $\lambda$.
(This can be accounted for using a different representation.) 
More importantly, for $\lambda \approx 1$ barostat effects naturally become
irrelevant and the intrinsic relaxation time $\tauA$ essentially sets $\thetaTT$ and not $\tstar \approx \Tgg$.
We thus observe $\thetaTT \approx \tauA \lambda^0 \kappa^0$ for all $\lambda > 0.8$ 
as seen in table~\ref{tab} for $\kappa=10^{-2}$.
Basically, since we have {\em two} time scales $\tauA$ and $\Tgg$,
it is not possible to put all $\thetaTT(\lambda,\kappa)$ on {\em one} master curve.

\section{Conclusion}
\label{sec_conc}

\paragraph*{Static properties.}
Focusing on two-dimensional elastic networks (sect.~\ref{sec_algo}) 
we have characterized theoretically (sect.~\ref{sec_theo}) and numerically (sect.~\ref{sec_simu}) 
the static and dynamical fluctuations of shear-strain and shear-stress in generalized Gaussian 
ensembles as a function of the dimensionless parameter $\lambda = \Gext/(\Geq + \Gext)$, 
fig.~\ref{fig_sketch}. 
Monitoring various properties we interpolate between the $\NVgT$-ensemble 
($\lambda=1$) and the $\NVtT$-ensemble ($\lambda=0$) and consider even negative $\lambda$. 
The static strain-strain fluctuations $\muGG$ (fig.~\ref{fig_dgam}), the strain-stress fluctuations $\muGT$ 
and the stress-stress fluctuations $\muTT$ (fig.~\ref{fig_muF_L}) have been shown to decrease linearly 
with increasing $\lambda$ \cite{foot_tildefunct}.
As a consequence, the static shear modulus $\Geq$ 
may be obtained (fig.~\ref{fig_muF_L}) from either the strain-strain fluctuation formula $\Ggg$, 
eq.~(\ref{eq_Ggggen}), or from the strain-stress formula $\Ggt$, eq.~(\ref{eq_Ggtgen}), for $\lambda < 1$ 
or the stress-stress fluctuation formula $\Gtt$, eq.~(\ref{eq_Gttgen}), for $\lambda \ne 0$. 
The latter formula is a generalization of the well-known stress-fluctuation formula eq.~(\ref{eq_Gtt}) 
for $\lambda = 1$. 
When sampled from a {\em finite} time series $(\gamhat_k,\tauhat_k)$, $\Gggt$, $\Ggtt$ and $\Gttt$
behave similarly with time $t$ for all $\lambda$ (fig.~\ref{fig_Geq_t}). 
Interestingly, $\Gttt|_{\lambda=1}$ and $\Ggtt|_{\lambda<1}$
are identical for {\em  all} times.

\paragraph*{Dynamic properties.}
The influence of the parameter $\kappa$ of the MC shear-barostat (sect.~\ref{algo_finiteT}) 
has been investigated for various dynamical properties, especially for the 
stress-stress correlation function $\cTTt$ (fig.~\ref{fig_cTTt_Gxymax}). 
The upper time limit $\tstar(\lambda)$, below which barostat effects can be ignored,
is set by the strain-strain crossover time $\Tgg(\lambda)$ (fig.~\ref{fig_gGGt}).
While a rapid barostat may be of advantage for static properties, 
a sufficiently slow barostat with a large $\tstar$ corresponds to the fundamentally important 
linear-response limit where the barostat is only relevant for the distribution of the start points 
of the trajectories (and, hence, $C_\mathrm{ab}(0) = \muAB$) but {\em not} for their dynamical 
pathways for $t \ll \tstar$. 
We have argued that this implies the fundamental scaling $\gABt \sim \lambda^0 \kappa^0$ 
for $t \ll \tstar(\lambda)$ as demonstrated explicitly for $\gTTt$ in fig.~\ref{fig_gTTt}.
Assuming such a slow barostat one may obtain $\Geq$
using eq.~(\ref{eq_keydyna}) from $\cTTtlam \approx (1- \lambda) \Geq$ for $\lambda < 1$ 
(fig.~\ref{fig_cTTt}). 
As emphasized elsewhere \cite{WXB15,WXBB15}, it is impossible, however,
to obtain the modulus $\Geq$ {\em alone} from the autocorrelation function $\cTTt|_{\lambda=1}$.
The experimentally important shear-stress relaxation modulus $G(t)$ may be most readily 
determined at $\lambda=1$ calculating first $\Geq = \muA - \muTT|_{\lambda=1}$
and then $G(t) = \cTTt|_{\lambda=1} + \Geq$ \cite{foot_oscil}. 
We have commented briefly on the creep compliance $J(t)$. Being well-described by
the Kelvin-Voigt model, eq.~(\ref{eq_KelvinVoigt}), $J(t)$ is best obtained from the 
strain-strain MSD $\gGGt$ at $\lambda=0$ (fig.~\ref{fig_gGGt}). 

\paragraph*{More general thermodynamic variables.}
Most of the theoretical relations and numerical techniques discussed above,
especially the key formulae eq.~(\ref{eq_keystat}) and eq.~(\ref{eq_keydyna}),
generalize readily for any pair of {\em continuous} extensive and intensive variables $X$ and $I$
with $\Meq = \partial I / \partial X$ being the equilibrium modulus. 
With $\Mext$ being the spring constant of the external potential controlling the fluctuations
of the extensive variable and $\lambda = \Mext/(\Meq + \Mext)$ one obtains, e.g.,
the relaxation modulus
\begin{equation}
M(t) = C(t)|_{\lambda=0} = \Ctlam + \lambda \Meq \ \mbox{ for } t \ll \tstar(\lambda) 
\label{eq_keydynagen}
\end{equation}
with $C(t) \equiv \beta V \langle \delta \Ihat(t) \delta \Ihat(0) \rangle$
being the relevant autocorrelation function. 
The associated MSD $g(t) \equiv (\beta V/2) \langle (\Ihat(t)-\Ihat(0))^2 \rangle$ 
is expected to be strictly $\lambda$-independent in the same time window.
These properties must be computed again either using a slow ``barostat" with a large, 
albeit finite $\tstar \ll \ttraj$ or, equivalently, by averaging over an equilibrium 
ensemble of configurations with frozen $\Xhat$. 
For times $t \gg \tstar$ a switched-on ``barostat" ultimately restores the ergodicity 
and $C(t)$ must vanish while $g(t)$ approaches its finite thermodynamic value 
$\beta V \langle \delta \Ihat^2 \rangle$.

\vspace*{0.2cm} 
\begin{acknowledgments}
H.~Xu thanks the IRTG Soft Matter for financial support.
We are indebted to O.~Benzerara (Strasbourg)
and J.~Helfferich (Freiburg) for helpful discussions.
\end{acknowledgments}


\begin{thebibliography}{36}
\bibitem[{foo({\natexlab{c}})}]{foot_tildefunct}
\bibinfo{note}{Using the slightly modified definitions
  $\tilde{\mu}_\mathrm{\gamma\gamma} = \muGG \Geq$,
  $\tilde{\mu}_\mathrm{\gamma\tau} = \muGT$ and $\tilde{\mu}_\mathrm{\tau\tau}
  = \muGT/\Geq$, one obtains the general transform
  $\tilde{\mu}_\mathrm{ab}|_{\lambda} = \tilde{\mu}_\mathrm{ab}|_{\lambda=0} -
  \lambda$ for all fluctuations. Similarly, the rescaling
  $\tilde{C}_\mathrm{\gamma\gamma}(t) = \cGGt \Geq$,
  $\tilde{C}_\mathrm{\gamma\tau}(t) = \cGTt$ and
  $\tilde{C}_\mathrm{\tau\tau}(t) = \cTTt/\Geq$ leads to the compact
  transformation relation $\tilde{C}_\mathrm{ab}(t)|_{\lambda} =
  \tilde{C}_\mathrm{ab}(t)|_{\lambda=0} - \lambda$ for the correlation
  functions. Equation~(\ref{eq_blurb}) may be generalized as $$
  \frac{\ddiff}{\ddiff \lambda} \tilde{\mu}_\mathrm{ab}|_{\lambda} =
  \frac{\ddiff}{\ddiff \lambda} \tilde{C}_\mathrm{ab}(t)|_{\lambda} = -1 $$
  with appropriate conditions at $\lambda=0$.}
\bibitem[{foo({\natexlab{d}})}]{foot_ergodic}
\bibinfo{note}{This assumes that the system is ergodic and does not apply if
  one samples, e.g., configuration ensembles with quenched shear-strain.}
\bibitem[{foo({\natexlab{e}})}]{foot_theta_f}
\bibinfo{note}{As seen in fig.~\ref{fig_thetaTT}, in some cases rather long
  trajectories are needed to determine clearly a relaxation time $\thetaAB$
  from the $1/t$-decay of $y(t) \equiv 1 - \muABt/\muAB$. As shown in
  ref.~\cite{WXBB15} one may instead characterize a relaxation time using $y(t
  = \theta_f) = f$ with $f \ll 1$ being a fixed fraction, say $f=1/100$
  \cite{WXBB15}. If $f$ is sufficiently small, eq.~(\ref{eq_muABt_asymp})
  implies $\thetaAB= \theta_f f/2$.}
\bibitem[{foo({\natexlab{f}})}]{foot_prime}
\bibinfo{note}{A prime denotes generally a derivative of a function with
  respect to its argument.}
\bibitem[{foo({\natexlab{h}})}]{foot_CtGauss}
\bibinfo{note}{A good fit for the short time behavior is given by the Gaussian
  $\cTTt = \muTT \exp(-(t/\tilde{\theta})^2/2)$ \cite{WXBB15}. To leading order
  this yields eq.~(\ref{eq_gTTt_determinist}).}
\end{thebibliography}

\end{document}